\documentclass[10pt,twocolumn,journal]{IEEEtran}
\usepackage{float}
\usepackage{latexsym}
\usepackage[hidelinks]{hyperref}
\usepackage{amsfonts}
\usepackage{amsbsy}
\usepackage{amssymb}
\usepackage{times}
\usepackage{graphicx}
\usepackage{setspace}
\usepackage{enumerate}
\usepackage[usenames]{color}
\usepackage[dvips]{pstcol}
\usepackage{epstopdf}
\usepackage[caption=false]{subfig}
\usepackage{cite}
\usepackage{amssymb}
\usepackage{amsfonts}
\usepackage{graphicx}
\usepackage{epsfig}
\usepackage{psfrag}
\usepackage{xcolor}
\usepackage{amsfonts, bm}
\usepackage{epstopdf}
\usepackage{cite}
\usepackage{color}
\usepackage{xcolor}
\usepackage{subfig}
\usepackage{verbatim}
\usepackage{multirow}
\usepackage{booktabs}
\usepackage{amsthm}
\usepackage{makecell}
\usepackage{units}
\usepackage[linesnumbered, ruled]{algorithm2e}
\usepackage{algpseudocode}
\usepackage{amsmath}

\usepackage[linesnumbered, ruled]{algorithm2e}
\usepackage{algpseudocode}
\usepackage{amsmath}

\IEEEoverridecommandlockouts

\begin{document}
	
	\title{From Analog to Digital: Multi-Order Digital Joint Coding-Modulation for Semantic Communication}
	\author{\IEEEauthorblockN{Guangyi Zhang, Pujing Yang, Yunlong Cai, Qiyu Hu, and Guanding Yu}
	\thanks{
	This work was supported in part by the 
	
	Guangyi Zhang, Pujing Yang, Qiyu Hu, and Guanding Yu  are with the College of Information Science and Electronic Engineering, Zhejiang University, Hangzhou 310027, China (e-mail: zhangguangyi@zju.edu.cn; yangpujing@zju.edu.cn; qiyhu@zju.edu.cn; yuguanding@zju.edu.cn). 
	
	Yunlong Cai is with the College of Information Science and Electronic Engineering, Zhejiang University, Hangzhou 310027, China, and also with the Peng Cheng Laboratory, Shenzhen 518071, China (e-mail: ylcai@zju.edu.cn).}} 
	\maketitle
	\vspace{-3.3em}

\begin{abstract}
	Recent studies in joint source-channel coding (JSCC) have fostered a fresh paradigm in end-to-end semantic communication. Despite notable performance achievements, present initiatives in building semantic communication systems primarily hinge on the transmission of continuous channel symbols, thus presenting challenges in compatibility with established digital systems. In this paper, we introduce a novel approach to address this challenge by developing a multi-order digital joint coding-modulation (MDJCM) scheme for semantic communications. Initially, we construct a digital semantic communication system by integrating a multi-order modulation/demodulation module into a nonlinear transform source-channel coding (NTSCC) framework. Recognizing the non-differentiable nature of modulation/demodulation, we propose a novel substitution training strategy. Herein, we treat modulation/demodulation as a constrained quantization process and introduce scaling operations alongside manually crafted noise to approximate this process.  As a result, employing this approximation in training semantic communication systems can be deployed in practical modulation/demodulation scenarios with superior performance. Additionally, we demonstrate the equivalence by analyzing the  involved  probability distribution. 
	Moreover, to further upgrade the performance, we develop a hierarchical dimension-reduction strategy to provide a gradual information extraction process. Extensive experimental evaluations demonstrate the superiority of our proposed method over existing digital and non-digital JSCC techniques.  
\end{abstract}
	
	
\IEEEpeerreviewmaketitle

\section{Introduction}
	The rapid evolution of sixth-generation (6G) communication systems has led to the emergence of numerous intelligent applications\cite{Weizhi_2023TCCN}, such as the Internet of Things (IoT), autonomous driving, and mobile Internet \cite{Saad_Netw2020}. In response to this shift, there has been a paradigm shift towards embracing semantic communications\cite{ Jialong_TCVT2022, Haotian_Arxiv2023,  Guangyi_TCOM2024,Xiaodong_IOTJ2024,Zhang_Sept2022}. Semantic communications hold significant promise for delivering superior transmission rate by extracting task-related information from raw data \cite{Zhenzi_2021, Jianhao_IOTJ2024,Wenyu_TWC2023, Cheng_TCOM2024, Danlan_JSAC2023,Tung_JSAC2022}.

\subsection{Prior Work}
	In contrast to the classical separation-based scheme using compression algorithms combined with the practical channel codes, semantic communications typically conceive of an integrated design of source codes and channel codes, also referred to as joint source-channel coding (JSCC) \cite{Huiqiang_TSP2021, Xinchao_Access2024,Jianhao_Arxiv2024, Dongxu_JCIN2023, Saidutta_JSAC2021, Mingyu_TCCN2022, Haotian_Arxiv, Lunan_JSAC2023, Shuaishuai_ICL2023, Xiaojiao_IOTJ2024}.  
	In particular, numerous researchers aimed to design JSCC systems utilizing deep learning techniques, predominantly focusing on leveraging autoencoder (AE) and its variants, such as variational AE (VAE). Several works have significantly enhanced the transmission efficiency across different data modalities.
	A pioneer work for text semantic communications has been proposed in \cite{Huiqiang_TSP2021}, demonstrating superior performance compared to conventional separation-based methods. 
	Another notable advancement is DeepJSCC, proposed in \cite{Eirina_TCCN2019}, which utilizes an AE-based framework to map the source images to low-dimensional feature vectors, effectively mitigating the cliff effect.  Building upon \cite{Eirina_TCCN2019}, researchers have endeavored to upgrade performance by introducing advanced designs. 
	For instance, the attention mechanism is employed in \cite{Haotian_Arxiv, Guangyi_TCCN} to augment the adaptability of DeepJSCC over MIMO channels. Moreover, in \cite{Yang_ICASSP2023}, authors designed a semantic communication system using the Swin Transformer, achieving improved performance across various image resolutions and channel conditions. 
	
	More recently, entropy model-based learned image compression (LIC) mechanisms \cite{balle_arxiv2018, Dailan_CVPR2021, Zhengxue_CVPR2020,Zongyu_ICML2021} have achieved notable success in source coding. These methods utilize an entropy model to estimate the probability distribution of the feature vector to be compressed, which directly relates to the final code length using entropy encoding. The whole training is modeled as a problem of rate-distortion (RD) optimization. 
	Inspired by these advancements, the authors of \cite{Jincheng_JSAC2022} introduced nonlinear transform source-channel coding (NTSCC) for semantic communications. 
	Taking into account that images vary in the amount of information they contain and necessitate different numbers of bits for compression, NTSCC integrates JSCC with an entropy model. This integration allows for the determination of optimal transmission bandwidth for each image, enabling variable-rate transmission and positioning NTSCC as a potent JSCC scheme for image transmission. Building upon NTSCC, an improved version termed NTSCC+ has been developed in \cite{Sixian_JSAC2023}. NTSCC+ introduces a context model to capture contextual information in the latent feature vector, thereby further enhancing the overall performance of semantic communication systems.

    Though deep learning-based JSCC systems have shown remarkable performance,  their practical deployment in modern digital communication devices encounters several challenges. 
   Many existing works primarily focus on directly mapping the source information to continuous channel symbols, a process that often necessitates analog transmission or a full-resolution constellation \cite{Peiwen_TCOM2022, Peiwen_JSAC_2023}. However, this approach poses challenges for power amplifiers with non-ideal characteristics, thereby limiting the feasibility and scalability of such systems in real-world applications.
    To tackle this challenge, researchers have explored various approaches, aiming to discretize the channel inputs to accommodate the hardware limitations \cite{Tze-Yang_ICC2022}. For instance, in \cite{Kristy_ICML2019}, the authors modeled latent variables as sequences of bits following a Bernoulli distribution. Within this framework, the encoder outputs the probability that the bit symbol is either $1$ or $0$, with exact channel symbols sampled from this distribution. This enables more efficient representation and transmission of information within hardware constraints.
    Similarly, an innovative end-to-end optimized JSCC scheme for wireless image transmission has been presented in \cite{Yufei_Arxiv2023}. In this scheme, the encoder outputs the probabilities associated with each constellation point, which are then sampled according to these probabilities. In \cite{Joohyuk_TCCN,Chuanhong_OFDM}, the authors proposed to transform the semantic features into discrete bits.  Nevertheless, the quantization operation entails mapping continuous variables to a finite number of discrete variables, posing a challenge for neural networks due to their non-differentiable nature.
    This renders common optimization methods, such as stochastic gradient descent (SGD), ineffective. To address this issue, existing digital communication systems employ various strategies to make the modulation process trainable. For instance, in \cite{Qiyu_TWC2023}, the authors considered the straight-through estimator (STE). Additionally, the Gumbel-Softmax strategy is employed as a continuous approximation of sampling \cite{Yufei_Arxiv2023}, enabling backpropagation through the modulation process.
    
\subsection{Motivation and Contributions}
    While current methods have advanced digital semantic communications, several challenges remain. First, many approaches rely on small datasets, like CIFAR10, which may not accurately reflect the complexities of real-world scenarios. Second, existing digital systems typically support fixed-rate transmissions and lack the flexibility to determine the optimal bandwidth for each source sample. This limitation can result in suboptimal system performance. Third, most current methods only support a fixed type of modulation, necessitating model updates when the modulation order changes, which hinders adaptability.
    To address these challenges, we propose a multi-order digital joint coding-modulation scheme tailored for wireless image transmission that addresses issues related to resolution generalization, and modulation flexibility. 
    To this end, we initiate the development of a modulation-agnostic digital semantic communication system by integrating a multi-order modulation/demodulation module into an NTSCC framework.  To tackle the non-differentiable characteristics of modulation/demodulation, we propose a novel substitution training strategy, demonstrating superior effectiveness compared to existing methods. Subsequently, we conduct a theoretical analysis to validate the rationale behind the proposed training strategy. To further enhance performance, we devise a hierarchical dimension-reduction strategy to facilitate a gradual information extraction process.
   
	Specifically, we summarize our main contributions as follows:
	\begin{itemize}
		\item[(1)] We introduce MDJCM for digital wireless image transmission, aiming to overcome the impracticality of training and utilizing multiple models for various modulation orders. To achieve this, we devise a multi-order modulator/demodulator, employing a conditional response network. This module considers the required modulation order and the input source, enabling MDJCM to generate distinct channel symbols based on the given order. Through optimization via training with diverse input sources and modulation orders, MDJCM can effectively determine the transmission bandwidth for each input source.
		
		\item[(2)] The requirement to map continuous variables to a finite set of symbols poses a significant challenge for model training. This process lacks differentiability, making conventional optimization algorithms like SGD ineffective. In this study, we tackle this challenge through the development of a novel substitution training strategy. Specifically, we treat the modulation/demodulation process as a constrained quantization operation and introduce scale operations and handcrafted noise as continuous relaxations to approximate this process. Crucially, this approximation is differentiable, allowing for its direct application to practical modulation processes. We further validate this approximation by comparing the probability distributions of the two processes, establishing their equivalence.
		
		\item[(3)] Recognizing the discrepancy between the modulation/demodulation process and the proposed continuous relaxation, we propose a two-phase strategy. Initially, we train with continuous relaxation and then fine-tune using STE to effectively address this discrepancy and enhance system performance. To further augment the performance of MDJCM, we introduce an innovative hierarchical dimension-reduction strategy at the transmitter. This strategy gradually reduces the intermediate feature dimension, directly corresponding to the number of transmitted channel symbols. Additionally, we implement a turbo decoder at the receiver, a technique known for its effectiveness in error correction \cite{Yihan_NIPS2019}.  

		\item[(4)] We conduct extensive simulations to validate the performance, encompassing images of varying resolutions.  Moreover, to demonstrate the effectiveness of each proposed scheme, we perform a comprehensive ablation study. Our experimental findings indicate that our proposed MDJCM achieves significant performance and remarkable compatibility, exhibiting superior performance compared to both emerging analog transmission schemes and digital transmission schemes.
	\end{itemize}

	\begin{figure*}[t]
		\begin{centering}
			\includegraphics[width=0.91 \textwidth]{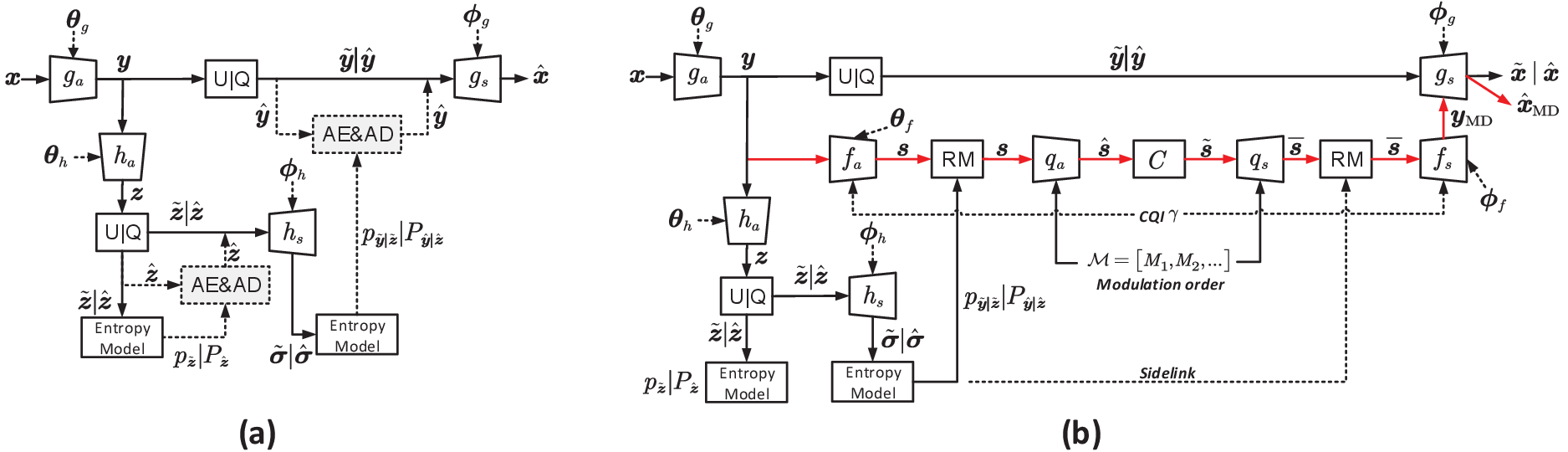}
			\par \end{centering}
		\caption{Operational diagrams. (a) The diagram of a typical learning-based source coder, \fbox{AE\&AD} represents arithmetic encoding and decoding, which are only employed in the compression phase. \fbox{U/Q} denotes adding uniform noise or quantization, where uniform noise is considered in the training phase and the quantization operation is employed for compression; (b) The diagram of the developed entropy model-based digital semantic communication system. \fbox{RM} denotes the rate matching operation, which is used to control the overall transmission overhead. \fbox{C} represents wireless channel.}
		\label{FrameArchitect}
	\end{figure*}

	The remainder of this paper is structured as follows. In Section \ref{SEC2}, we provide an overview of the proposed MDJCM. The design of the substitution training strategy is detailed and introduced in Section \ref{SEC3}. The detailed model architectures and the proposed hierarchical dimension-reduction strategy are presented in Section \ref{SEC4}. The simulations are presented in Section \ref{SEC5}. Finally, the concluding remarks are provided in Section \ref{SEC6}.
	
	\emph{Notations:}  For a vector $\mathbf{a}$, $\|\mathbf{a}\|$ is its Euclidean norm. 
	$p_x$ denotes a probability density function (PDF) of the continuous-valued random variable $x$, while $P_{\hat{x}}$ represents the probability mass function (PMF) of the discrete-valued random variable $\hat{x}$.
	Additionally, we denote $\mathcal{U}(-d,d)$ as the uniform distribution with a range from $-d$ to $d$.
	Finally, ${\mathbb{C}^{m \times n}}({\mathbb{R}^{m \times n}})$ are the space of ${m \times n}$ complex (real) matrices.

	\section{Problem Formulation and System Overview}\label{SEC2}
	In this section, our objective is to provide an overview of the proposed MDJCM. We commence by introducing the learning-based source coder, which encompasses variational entropy modeling, as well as analysis and synthesis transforms. 
	Subsequently, we present the MDJCM, encompassing learning-based source coder, encoding/decoding, and modulation/demodulation phases. 
	Notably, the learning-based source coder serves as a rate indicator, guiding the MDJCM in selecting an appropriate rate for transmission.
	
	\subsection{Learning-Based Source Coder}
	 Recently, a number of learning-based source coders have been designed to surpass the performance of many traditional coders\cite{Jinming_CVPR2023}, such as versatile video coding (VVC) and better portable graphics (BPG). The success of these methods stems from advanced entropy modeling, coupled with an autoencoder architecture for nonlinear transform coding. These approaches focus on minimizing the entropy of the encoded feature representations, approximated by a well-established entropy model. Consequently, the length of the final codeword can be optimized.

	 A typical learning-based source coder is depicted in Fig. \ref{FrameArchitect}(a). Given a raw data sample, such as an image represented by a pixel intensity vector $\bm{x} \in \mathbb{R}^{n_x}$\footnote{We use the notation ``$n_x$" to denote the vector length, with subscripts indicating the specific vector.}, a deep learning-based nonlinear analytic transform $g_a$ is employed to convert $\bm{x}$ into a latent representation $\bm{y}=g_a(\bm{x}; \bm{\theta}_g)\in \mathbb{R}^{n_y}$, where $\bm{\theta}_g$ denotes the trainable parameters. Subsequently, $\bm{y}$ undergoes quantization to obtain a discrete latent representation $\bm{\hat{y}}=Q(\bm{y})\in \mathbb{R}^{n_y}$, with $Q(\cdot)$ representing the quantization operation. \color{black}At this stage, lossless compression of $\bm{\hat{y}}$ into a bit sequence is feasible via entropy coding methods like arithmetic coding or range coding, given that the entropy information (model) $P_{\bm{\hat{y}}}$ is known. 
	 In particular, 
	 we further define the rate $R$ as the expected length of the bit sequence, which is equivalent to the entropy of $\bm{\hat{y}}$, where $R$ can be specifically determined by the entropy model as follows:
	 \begin{equation}
	 	R=\mathbb{E}_{\bm{x} \sim p_{\bm{x}}}\left[-\log _2 P_{\bm{\hat{y}}}\left(Q\left(g_a\left(\bm{x} ; \bm{\theta}_g\right)\right)\right)\right].
	 \end{equation}
	 Following this, the bit sequence then undergoes entropy decoding for lossless reconstruction of $\bm{\hat{y}}$. Subsequently, a nonlinear synthetic transform $g_s$ is utilized to reconstruct the decoded data $\bm{\hat{x}}=g_s(\bm{\hat{y}};\bm{\phi}_g)$, where $\bm{\phi}_g$ encompasses its trainable parameters.
	 
	 As stated above, the compression of the quantized discrete latent $\bm{\hat{y}}$ necessitates its entropy information (model). One of the most effective methods to estimate the entropy is the hyperprior model proposed in \cite{balle_arxiv2018}. As shown in Fig. \ref{FrameArchitect}(a), the main idea of the hyperprior model is to introduce another latent feature $\bm{z}=h_a(\bm{y};\bm{\theta}_h)\in \mathbb{R}^{n_z}$, where $h_a$ represents the parametric synthesis transform and $\bm{\theta}_h$ denotes the trainable parameters. 
	 Then, $\bm{\hat{z}}=Q(\bm{z})\in \mathbb{R}^{n_z}$ is taken as the side information, serving to characterize the dependencies among the elements in $\bm{\hat{y}}$. 
	 In general, the optimization of such a learning-based coder involves a tradeoff between rate and distortion, which can be written as 
	 \begin{equation}\label{loss_func}
	 	\begin{aligned}
	 		R+\lambda \cdot D & =\mathbb{E}_{\bm{x} \sim p_{\bm{x}}}\big[\underbrace{-\log _2 P_{\bm{\hat{y}} | \bm{\hat{z}}}(\bm{\hat{y}} | \bm{\hat{z}})}_{\text{compression rate}}\underbrace{-\log _2 P_{\bm{\hat{z}}}(\bm{\hat{z}})}_{\text{side info rate}}\big] \\
	 		& +\lambda \cdot \mathbb{E}_{\boldsymbol{x} \sim p_{\bm{x}}}[D(\bm{x}, \bm{\hat{x}})],
	 	\end{aligned}
	 \end{equation}
	 where $D(\cdot)$ denotes the distortion function, and $\lambda$ controls the tradeoff between the distortion and compression rate. Specifically, with a higher $\lambda$, the resulting model will compress the source at a higher rate.  
	 
	 Due to the non-differentiability of the quantization operation during training,  the quantization operations on $\bm{y}$ and $\bm{z}$ are  replaced by the addition of  uniform noise, represented by $\bm{\tilde{y}} = \bm{y} + \bm{u} \in \mathbb{R}^{n_y}$ and $\bm{\tilde{z}} = \bm{z} + \bm{u}\in \mathbb{R}^{n_z}$, respectively, where $\bm{u}$ is randomly sampled from standard uniform distribution $\mathcal{U}(-\frac{1}{2},\frac{1}{2})$. Then, each $\tilde{y}_i$ in $\bm{\tilde{y}}$ is variationally modeled as a Gaussian distribution, where the standard deviation $\sigma_i$ and mean $\mu_i$ are predicted based on $\bm{\tilde{z}}$, i.e.,
	 \begin{equation}
	 	p_{\bm{\tilde{y}} | \bm{\tilde{z}}}\left({\tilde{y}}_i | \bm{\tilde{z}}\right)=\left(\mathcal{N}(\!\tilde{\mu}_i, \tilde{\sigma}_i^2) * \mathcal{U}(-\frac{1}{2}, \frac{1}{2})\right) \left(\tilde{y}_i\right),
	 \end{equation}
 	\begin{equation}
 		p_{\bm{\tilde{y}} \mid \bm{\bm{\tilde{z}}}}(\bm{\tilde{y}} | \bm{\tilde{z}})=\prod_i p_{\bm{\tilde{y}} \mid \bm{\tilde{z}}}\left(\tilde{y}_i | \bm{\tilde{z}}\right),
 	\end{equation}
	 where $(\bm{\tilde{\mu}}, \bm{\tilde{\sigma}})=h_s\left(\bm{\tilde{z}} ; \bm{\phi}_h\right)$, with its $i$-th element being $(\tilde{\mu}_i, \tilde{\sigma}_i)$, $\bm{\phi}_h$ denotes the trainable parameters of the parametric analysis transform $h_s$, and $*$ represents the convolutional operation. 
	 For the term of side information rate, $-\log _2 p_{\bm{\hat{z}}}(\bm{\hat{z}})$, we do not have prior beliefs. Thus, it is modeled as the non-parametric fully factorized density, as given by
	 \begin{equation}
	 	p_{\bm{\tilde{z}} | \bm{\psi}}(\bm{\tilde{z}} | \bm{\psi})=\prod_i\left(p_{z_i | \bm{\psi}^{(i)}}(\bm{\psi}^{(i)}) * \mathcal{U}(-\frac{1}{2}, \frac{1}{2})\right)\left(\tilde{z}_i\right),
	 \end{equation}
 	where $\bm{\psi}$ denotes the trainable parameters.
	 With these modelings, (\ref{loss_func}) can be employed as the loss function for learning the source coder. More specifically, the entropy term can be readily calculated by using this cumulative distribution function\cite{balle_arxiv2018}. By optimizing the rate term with varying  weights, the model is able to generate the latent representations $\bm{y}$ with different  levels of entropy (sparsity).

	\subsection{Formulation of Rate-Guided Digital JSCC}
	Inspired by \cite{Jincheng_JSAC2022}, we propose a multi-order digital semantic communication system, where the rate control is under the guidance of the learning-based source coder.
	Recalling that the compression rate at the source coder is correlated to the entropy model $P_{\bm{\hat{y}}}$, we are able to control the transmission bandwidth based on $P_{\bm{\hat{y}}}$. 
	The workflow of the proposed MDJCM is shown in Fig. \ref{FrameArchitect}(b), comprising two parts, the learning-based source coder and the digital JSCC.
	
	The digital JSCC consists of a joint source-channel (JSC) encoder $f_a$, a JSC decoder $f_s$, a modulator $q_a$ and a demodulator $q_s$, with the parameters of $f_a$ and $f_s$ being $\bm{\theta}_f$ and $\bm{\phi}_f$, respectively.  
	In general, wireless transmission systems should be able to overcome channel variations, e.g., fading effects. 
	To accommodate varying channel conditions, it may be necessary to train multiple networks or retrain a network as the channel conditions fluctuate. However, training and deploying multiple networks are not always practical, and thus we aim to design a digital semantic communication system that is compatible with fading, noise, etc. That is, MDJCM is expected to accommodate various channel signal-to-noise ratios (SNRs). Moreover, the same problems will appear when MDJCM is required to support different modulation orders. Therefore, MDJCM is designed to generate different orders of constellation points within one model. In the following, we will provide an overview of the proposed MDJCM.
	
	To this end, we consider the conditional coding design for the digital JSCC part. In particular, the JSC encoder takes the output of the analytic transform $\bm{y}$ and channel SNR (evaluated by a channel quality indicator (CQI) $\gamma$)  as input, and the encoding function can be expressed as 
	\begin{equation}
		\bm{s}=f_a(\bm{y};\gamma, \bm{\theta}_f) \in \mathbb{C}^{n_s}, 
	\end{equation}
	where $\bm{s}$ represents the channel input and $n_s$ denotes the number of transmitted channel symbols. 
	Different from existing semantic communication systems, we consider a digital transmission case. We specifically design a varying-order digital modulator as well as a demodulator, with the optional order set being defined as $\mathcal{M}=[M_1,M_2,\ldots]$. Then, the channel input $\bm{s}$ is further mapped to the discrete constellation symbols, as $\bm{\hat{s}}=q_a(\bm{s};M) \in \mathbb{C}^{n_s}$, where the modulator is able to generate constellation symbols in different orders. Herein, the channel bandwidth ratio (CBR) is defined as  $\rho = n_s/n_x$. Drawing inspiration from the benefits of encoding different source data with different rates \cite{balle_arxiv2018}, the entropy model is employed to indicate the CBR value for each source data. Specifically, if the entropy for certain source data is relatively large, a higher CBR value will be adopted. Typically, $\bm{s}$ is comprised of multiple embedding vectors $y_i$, each is length $C$. After the rate matching operation, the length of each vector is adjusted to ${k}_i=Q\left(-\eta \log P_{\hat{y}_i | \boldsymbol{\hat{z}}}\left(\hat{y}_i | \boldsymbol{\hat{z}}\right)\right)$, where $Q$ denotes the quantization operation, and $\eta$ controls the relation between the prior $P_{\hat{y}_i | \boldsymbol{\hat{z}}}\left(\hat{y}_i | \boldsymbol{\hat{z}}\right)$ and the expected the length of the corresponding symbol vector $s_i$.  The way to realize this will be depicted in Section IV-B.
	
	Subsequently, the encoded $\bm{\hat{s}}$ merely undergoes a power constraint $\mathcal{E}_s$, as given by $\| \bm{\hat{s}}  \|^2 \leq \mathcal{E}_s$. Then, $\bm{\hat{s}}$ is sent from the transmitter to the receiver through a wireless channel, which is given by 
	\begin{equation}
		\bm{\tilde{s}} = h \bm{\hat{s}} + \bm{n} \in \mathbb{C}^{n_s},
	\end{equation} 
	where $h$ denotes the channel gain coefficient, and $\bm{n}$ represents the additive white Gaussian noise (AWGN). In addition, $\mathbf{n} \sim \mathcal{N}\left(0, \sigma^2 \mathbf{I}\right)$, where $\sigma^2$ denotes the noise variance. To characterize the quality of the physical channel, we define the SNR as follows:
	\begin{equation}
	 \textrm{SNR}	=10 \log _{10} \frac{\mathcal{E}_s}{\sigma^2}.
	\end{equation} 
	At the receiver, the received noisy signal $\bm{\tilde{s}}$ is firstly demodulated to $\bm{\bar{s}}=q_s(\bm{\tilde{s}}; M) \in \mathbb{C}^{n_s}$. Then, $\bm{\bar{s}}$ undergoes decoding by $f_s$, i.e., $\bm{y}_{\text{MD}}=f_s(\bm{\bar{s}};\gamma,\bm{\phi}_f) \in \mathbb{R}^{n_y}$. Finally, the recovered source $\bm{\hat{x}}_{\text{MD}}$ is given by $\bm{\hat{x}}_{\text{MD}} = g_s(\bm{{y}}_{\text{MD}}; \bm{\phi}_g) \in \mathbb{R}^{n_x}$.  
	The target of the JSCC system is to find the optimal parameters $\left(\bm{\theta}_g^*,\bm{\theta}_h^*,\bm{\theta}_f^*, \bm{\phi}_f^*,\bm{\phi}_h^*,\bm{\phi}_g^* \right)$ to minimize the reconstruction distortion between the source data and the reconstructed data, as well as to find a suitable CBR value for transmitting each $\bm{x}$.  The training objective will be presented in the next section.

\section{Training with Substitution Process}\label{SEC3}
	In this section, our focus lies on the training strategies for MDJCM. When training the MDJCM, a significant issue arises due to the non-differentiability of the modulation and demodulation processes. This interrupts gradient propagation during MDJCM training. To address this challenge, we propose a novel mechanism involving training MDJCM with a differentiable substitution process. Consequently, the obtained model can be directly applied to digital transmission. In the following, we start by formulating the problem, followed by elaborating on the rationality of the proposed method. Subsequently, we propose a two-phase training strategy to further improve the system performance by eliminating the mismatch between the training phase and the testing phase.
	
\subsection{Problem Formulation}
	In the realm of learning-based compression, one of the key challenges is quantization. This is because the gradient of quantization is zero almost everywhere, making the gradient-based optimization unavailable. A classical solution to this is training the model with additive uniform noise to approximate quantization during the testing phase since training with additive uniform noise is differentiable. It has been demonstrated to be effective in various fields, achieving significant performance gains compared to other methods. This approach is particularly effective in enhancing the representation ability of the latent feature \cite{balle_arxiv2018,Dailan_CVPR2021,Zhengxue_CVPR2020,Zongyu_ICML2021}. 
	Drawing inspiration from this, we observe the modulation/demodulation operation is quite similar to the quantization, or rather, it can be viewed as a limited and scaled quantization process. Therefore, we propose to introduce a noise-adding process to approximate the modulation/demodulation operation during testing. To this end, we have the distribution constraint and process formulation as follows.
	
		\begin{figure}[t]
		\begin{centering}
			\includegraphics[width=0.28 \textwidth]{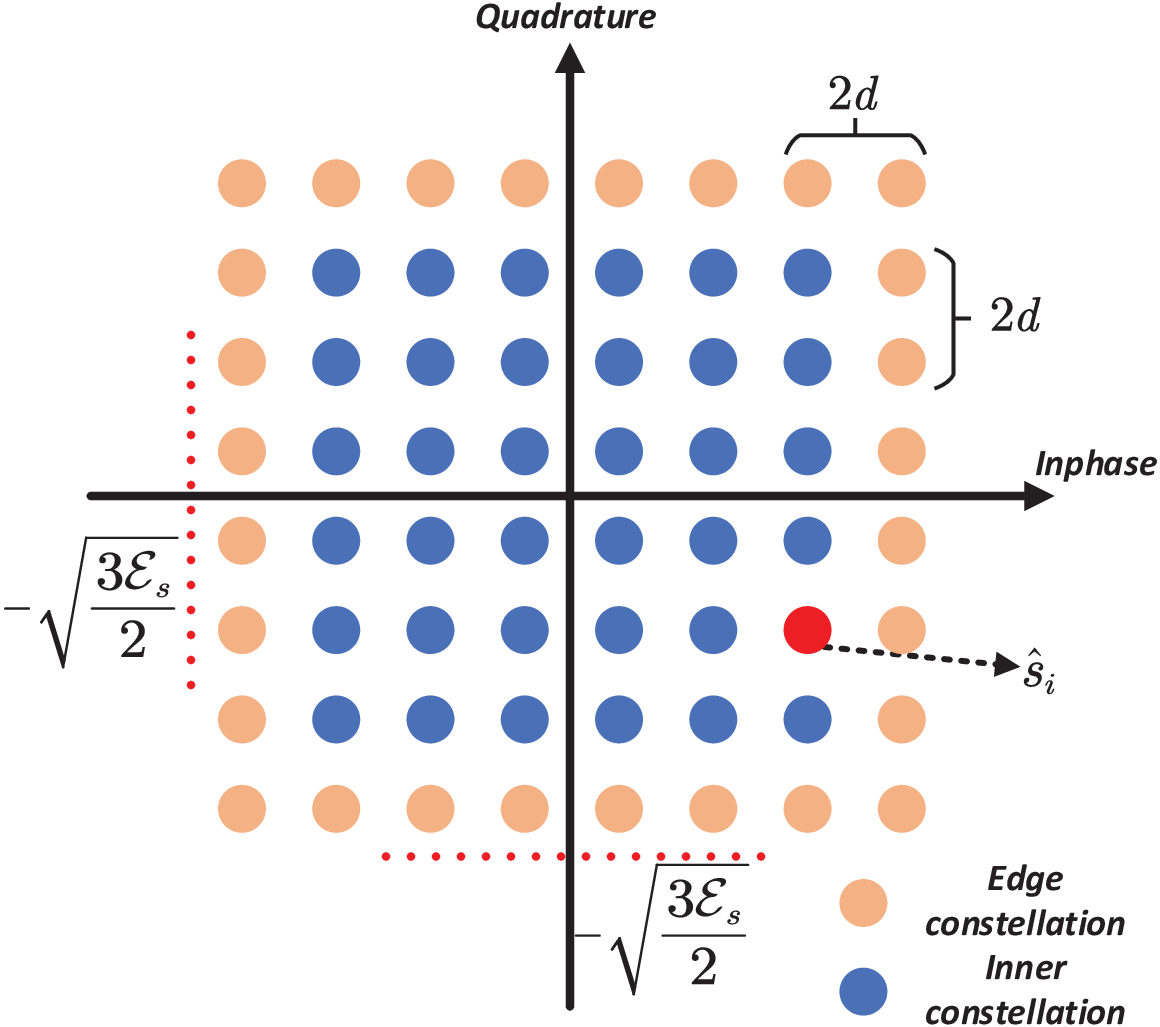}
			\par \end{centering}
		\caption{An example of $64$QAM constellation symbols. }
		\label{Modulation}
	\end{figure}
	
	\subsubsection{Distribution Constraint}
	We consider a more general case of quadrature amplitude modulation (QAM), as shown in Fig. \ref{Modulation}. There are in total of $M$ constellation points, the minimal distance between two constellations is $2d$. The average symbol power constraint is set to $\mathcal{E}_s$. 
	We denote each point as $(q^{In}_i, q^{Qu}_j)$, for $i,j=1,...,\sqrt{M}$, where $q^{In}_i$ and $q^{Qu}_i$ denote the  in-phase component and quadrature component, respectively. 
	As we aim to modulate $\bm{s}$ according to the constellation mapping, we constrain the power of $\bm{s}$ to improve the training stability. It is also of significance in ensuring the effectiveness, which we will present in Section \ref{SEC3B}. In particular, we make an assumption that the probability distribution of each element in $\bm{s}$ should satisfy that $p_{s_i}(u)= 0$, when $u <  -(\sqrt{M}-1)d$ and $u>  (\sqrt{M}-1)d$. 
	This constraint can be achieved by applying a clipping operation to the output of the rate matching operation.

\subsubsection{Process Formulation}
	In general, a typical digital communication system can be represented by the following process,
	\begin{equation}\label{process}
		\bm{s} \xrightarrow[\textit{}]{\text{Modulation}} \bm{\hat{s}}  \xrightarrow[\textit{}]{\times h/+\mathbf{n} \sim \mathcal{CN}\left(0, \sigma^2 \bm{I}\right)} \bm{\tilde{s}} \xrightarrow[\textit{}]{\text{Demodulation}} \bm{\bar{s}}.
	\end{equation}
		Since (\ref{process}) cannot be directly applied during the training phase, we aim to find an equivalent process for training. Considering the distribution constraint, we propose training the model using the following process:
	\begin{equation}\label{proprocess}
		\bm{s} \xrightarrow[\textit{}]{+\bm{u}_1 \sim \mathcal{U}(-d,d)} \bm{\hat{o}} \xrightarrow[\textit{}]{\times h/+\bm{n} \sim \mathcal{CN}(0,\sigma^2 \bm{I})} \bm{\tilde{o}} \xrightarrow[\textit{}]{+\bm{u}_2 \sim \mathcal{U}(-d,d)} \bm{\bar{o}},
	\end{equation}
	where we approximate the modulation by using scaled uniform noise, allowing (\ref{proprocess}) to serve as a continuous relaxation of (\ref{process}).   
	As stated in \cite{Gray_TIT1998}, the moments of the quantized random variable equal to the moments of the signal plus an independent random noise uniformly distributed on $(-d,d,]$, where $d$ denotes the quantization cell width. This indicates that it is rational to approximate the distortion with uniform noise caused by the modulation and demodulation process.
\subsection{Equivalence Illustration}\label{SEC3B}

	
	We next make an attempt to investigate the relationship between the distribution functions of $\bm{\bar{o}}$ and $\bm{\bar{s}}$, in order to examine whether process
	(\ref{proprocess}) can be considered as a continuous relaxation for process (\ref{process}).
	Specifically, the optimization objective of MDJCM can be formulated as \cite{Jincheng_JSAC2022}:
	\begin{equation}\label{eq16}
		\begin{aligned}
			& \min _{ } \mathbb{E}_{\bm{x} \sim p_{\bm{x}}} D_{\mathrm{KL}}\left[q_{\bm{\bar{o}}, \bm{\tilde{z}} | \bm{x}} \| p_{\bm{\bar{o}}, \bm{\tilde{z}} | \bm{x}}\right] \\
			  \leq &\min _{ } \mathbb{E}_{\bm{x} \sim p_{\bm{x}}} \mathbb{E}_{\bm{\bar{o}}, \bm{\tilde{z}} \sim q_{\bm{\bar{o}}, \bm{\tilde{z}} | \bm{x}}}\big[\log_2 q_{\bm{\bar{o}}, \bm{\tilde{z}} |\bm{x}}(\bm{\bar{o}}, \bm{\tilde{z}} | \bm{x})- \log_2 p_{\bm{\tilde{z}}}(\bm{\tilde{z}})-\\
			  &\mathbb{E}_{\bm{y} \sim p_{\bm{y} | \bm{\bar{o}} , \bm{\tilde{z}}}}\left[\log_2 p_{\bm{x} | \bm{y}}(\bm{x} |  \bm{y})\right] \big]- \log_2 p_{\bm{\bar{o}} | \tilde{\bm{z}}}(\bm{\bar{o}} | \bm{\tilde{z}})+\text {const} \\
		\end{aligned}
	\end{equation}
	The first term $q_{\bm{\bar{o}}, \bm{\tilde{z}} |\bm{x}}(\bm{\bar{o}}, \bm{\tilde{z}} | \bm{x})$ can be dropped since it will not introduce gradient the model parameters. The second term $p_{\bm{\tilde{z}}}(\bm{\tilde{z}})$ stands for the overhead of encoding the side information. The third term denotes the weighted distortion. Thus, we have the resulted learning objective for MDJCM:
 \begin{equation}\label{}
			\begin{aligned}
					L \triangleq R+\lambda D= &\mathbb{E}_{\bm{x} \sim p_{\bm{x}}}\big[\underbrace{-\log _2 p_{\bm{\bar{o}} | \tilde{\bm{z}}}(\bm{\bar{o}} | \bm{\tilde{z}})}_{\text{compression rate}}\underbrace{-\log _2 p_{\bm{\tilde{z}}}(\bm{\tilde{z}})}_{\text{side info rate}} \\
					& + \lambda [D(\bm{x}, \bm{\tilde{x}}_{\text{MD}})+ D(\bm{x}, \bm{\tilde{x}})]\big],
				\end{aligned}
		\end{equation}
	where $\lambda$ denotes the hyperparameter to control the tradeoff between rate and distortion.
	In \cite{Jincheng_JSAC2022}, the conditional entropy model $p_{\bm{\tilde{y}} | \bm{\tilde{z}}}(\bm{\tilde{y}} | \bm{\tilde{z}})$ is employed as a substitution of the fourth rate term $p_{\bm{\bar{o}} | \bm{\tilde{z}}}(\bm{\bar{o}} | \bm{\tilde{z}})$, since $p_{\bm{\tilde{y}} | \bm{\tilde{z}}}(\bm{\tilde{y}} | \bm{\tilde{z}})$  is proportional to $p_{\bm{\bar{o}} | \bm{\tilde{z}}}(\bm{\bar{o}} | \bm{\tilde{z}})$. In this case, the training objective is formulated as 
	 \begin{equation}\label{loss_func2}
		\begin{aligned}
			L \triangleq R+\lambda D= &\mathbb{E}_{\bm{x} \sim p_{\bm{x}}}\big[\underbrace{-\log _2 p_{\bm{\tilde{y}} | \tilde{\bm{z}}}(\bm{\tilde{y}} | \bm{\tilde{z}})}_{\text{compression rate}}\underbrace{-\log _2 p_{\bm{\tilde{z}}}(\bm{\tilde{z}})}_{\text{side info rate}} \\
			& + \lambda [D(\bm{x}, \bm{\tilde{x}}_{\text{MD}})+ D(\bm{x}, \bm{\tilde{x}})]\big],
		\end{aligned}
	\end{equation}
	However, the relation between $p_{\bm{\bar{o}} | \bm{\tilde{z}}}(\bm{\bar{o}} | \bm{\tilde{z}})$ and $P_{\bm{\bar{s}} | \bm{\tilde{z}}}(\bm{\bar{s}} | \bm{\tilde{z}})$ is still unknown, making it hard to deploy a model that is trained with (\ref{loss_func2}) to the practical modulation/demodulation scenarios. Fortunately, we find the entropy of $\bm{\bar{o}}$ serves as a continuous relaxation of the discrete entropy of $\bm{\bar{o}}$, under the distribution constraint. 
	
	Noticing that $\bm{\hat{o}}$ and $\bm{\hat{s}}$ are both complex vectors, we assume that the elements of the vector are independent from each other. For simplicity, we specifically analyze the $i$-th element of the vectors in processes (\ref{process}) and (\ref{proprocess}), where we use the lowercase letter to denote the element, e.g., $(\hat{o}_i^{In},\hat{o}_i^{Qu} )$ represents the $i$-th element of complex vector $\bm{\hat{o}}$, where the superscripts `Qu' and `In' denote the quadrature component and in-phase component, respectively. 
	In this case, we have 
	\begin{equation}
		\begin{aligned}
		&P(\bar{s}_i=((2m\!-\!\sqrt{M}\!-\!1)d,(2l\!-\!\sqrt{M}\!-\!1)d)) =\\
		&P(\bar{s}_{i}^{In}=(2m\!-\!\sqrt{M}\!-\!1)d)\times P(\bar{s}_{i}^{Qu}=(2l\!-\!\sqrt{M}\!-\!1)d),
		\end{aligned}
	\end{equation}
	where we omit the  conditional notation ``$|\bm{\tilde{z}}$" for simplicity.
	Furthermore, supposing that the in-phase and quadrature components are independent, we  will concentrate on the in-phase component in the following.
	As shown in Fig. \ref{Modulation}, we split the constellation symbols into two groups, the inner  constellation symbols and the edge constellation symbols. We will study these two types respectively, since the difference in calculating the probability distribution.
	
\subsubsection{Inner Constellation Symbols} 
	
	We concentrate on the in-phase component of the constellation symbol $((2m\!-\!\sqrt{M}\!-\!1)d,(2l\!-\!\sqrt{M}\!-\!1)d)$, i.e., $(2m\!-\!\sqrt{M}\!-\!1)d$. Moreover, we define the  transmission error as $\bm{e}=\bm{\tilde{s}}-\bm{\hat{s}}$ or $\bm{e}=\bm{\tilde{o}}-\bm{\hat{o}}$, where we assume the error is independent of the signal. It is important to note that this is available in many transmission models. In this case, the PDF of the $i$-th in-phase components of $\bm{e}$  is represented by $p_{e_i^{In}}$. Intuitively, this error is primarily influenced by the channel environments.  The probability of receiving $(2m\!-\!\sqrt{M}\!-\!1)d$ with process (\ref{process}) can be written as
	\begin{equation}\label{Pdenisty}
		\begin{aligned}
			& P(\bar{s}_i^{In}=(2m\!-\!\sqrt{M}\!-\!1)d)\\
			& =\sum_{t=1}^{\sqrt{M}}P(\hat{s}_i^{In}=(2t\!-\!\sqrt{M}\!-\!1)d)) \\
			 & \quad \times P(\bar{s}_i^{In} =(2m\!-\!\sqrt{M}\!-\!1)d|\hat{s}_i^{In}=(2t\!-\!\sqrt{M}\!-\!1)d)  \\
			 &=\sum_{t=1}^{\sqrt{M}} \big(\!\int_{(2t\!-\!2\!-\!\sqrt{M})d}^{(2t\!-\!\sqrt{M})d}p_{s^{In}_i}(\mu) d\mu  \! \int_{2(m-t)d-d}^{2(m-t)+d}p_{e_i^{In}}(\tau)d\tau \big).
		\end{aligned}
	\end{equation}

	Moreover, the probability of $\bar{o}_i^{In}$ equaling $(2m\!-\!\sqrt{M}\!-\!1)d$ can be represented by
	\begin{equation}\label{denisty}
	\begin{aligned}
		& p_{\bar{o}_i^{In}}((2m\!-\!\sqrt{M}\!-\!1)d)=  \\
		&  \int_{-\sqrt{M}d}^{\sqrt{M}d} \! \int^{\nu+d}_{\nu-d}\!p_{s^{In}_i}\!(\mu) d\mu\! \int_{(2m\!-\!\sqrt{M}\!-\!2)d-\nu}^{(2m\!-\!\sqrt{M}\!)d-\nu} \!
		p_{e_i^{In}}(\tau) \frac{1}{4d^2}  d\tau d \nu. 
	\end{aligned}
	\end{equation}
	where the detailed derivations can be found in Appendix A.  
	By comparing (\ref{Pdenisty}) with (\ref{denisty}), we further find that $ P(\bar{s}_i^{In}=(2m\!-\!\sqrt{M}\!-\!1)d) \approx  p_{\bar{o}_i^{In}}((2m\!-\!\sqrt{M}\!-\!1)d) * 2d$ indicating that process (\ref{process}) serves as an approximation of (\ref{proprocess}). When the modulation order gets larger, this approximation is much more accurate. More specifically, $p_{\bar{o}_i^{In}}$ provides a continuous relaxation by interpolating the discrete probability of ${\bar{s}_i^{In}}$ at the positions of the discrete constellation symbols. Most importantly, we observe that this equivalence is not limited by the channel environment, demonstrating that our proposed substitution training method offers strong generalizationability.

\subsubsection{Edge Constellation Symbols} 
	Similar to (\ref{Pdenisty}) and (\ref{denisty}), we can obtain probability (density) of receiving the edge constellation symbols, e.g., $(\sqrt{M}-1)d$, as given by
	\begin{equation}
	\begin{aligned}
		& P(\bar{s}_i^{In}=(\sqrt{M}-1)d)=\\
		&\sum_{t=1}^{\sqrt{M}} \left(\int_{(2t\!-\!2\!-\!\sqrt{M})d}^{(2t\!-\!\sqrt{M})d}p_{s_i^{In}}(\mu) d\mu  \int^{\infty}_{(1-2t)d}p_{e_i^{In}}(\tau) d\tau \right).
	\end{aligned}
	\end{equation}
	\begin{equation}
	\begin{aligned}
		& p_{\bar{o}_i^{In}}((2m\!-\!\sqrt{M}\!-\!1)d) =\\
		&\int_{-\sqrt{M}d}^{\sqrt{M}d} \int^{\nu+d}_{\nu-d}p_{s_i^{In}}(\mu)  d\mu \int_{(\!\sqrt{M}\!-\!2)d-\nu}^{\sqrt{M}d-\nu} p_{e_i^{In}}(\tau)  d\tau d \nu . 
	\end{aligned}
	\end{equation}
	Although ensuring equivalence becomes increasingly challenging for edge points, we find that this approximation remains accurate, particularly in higher SNR regimes, as demonstrated by simulations. Moreover, we further address this with the proposed two-phase training strategy. 
	
\subsection{Two-Phase Training Strategy} 
	This noise-adding strategy proves advantageous for enhancing representation ability, unlike the popular STE \cite{Qiyu_TWC2023} and soft-to-hard annealing\cite{Yufei_Arxiv2023, Tze-Yang_ICC2022}, which encounter training issues such as biased or unstable gradients, resulting in suboptimal models. While process (\ref{proprocess}) provides a continuous relaxation, enabling end-to-end training of the system, we identify two challenges in directly employing (\ref{proprocess}),
	\begin{itemize}
		\item There exists a mismatch between the training phase and the testing phase. Specifically, the modulation and demodulation operations are deterministic processes with respect to the signal, rather than truly random noise. Thus training with the noise-adding process (\ref{proprocess}) may hurt the system performance.
		
		\item Due to the constraints imposed by limited modulation order and channel noise, utilizing process (\ref{proprocess}) may not yield sufficient accuracy, especially when the modulation order is small. Additionally, there exists a bias in the approximation of edge constellation points.
	\end{itemize} 

To tackle this issue, we propose a two-phase training strategy. Initially, we train the entire model using process (\ref{proprocess}). This helps in obtaining a robust transmitter model with excellent representation capability. Subsequently, we freeze the parameters at the transmitter, $\left(\bm{\theta}_g,\bm{\theta}_h,\bm{\theta}_f \right)$, thereby retaining the encoder's ability. Next, we fine-tune the parameters at the receiver, $\left( \bm{\phi}_f,\bm{\phi}_h,\bm{\phi}_g \right)$, based on STE, aiming to alleviate discrepancies between the training and testing phases.  This is because STE operates as a deterministic function and does not introduce randomness. Our two-phase strategy simultaneously combines the benefits of both methods, resulting in enhanced performance.

	\begin{figure}[t]
	\begin{centering}
		\includegraphics[width=0.48 \textwidth]{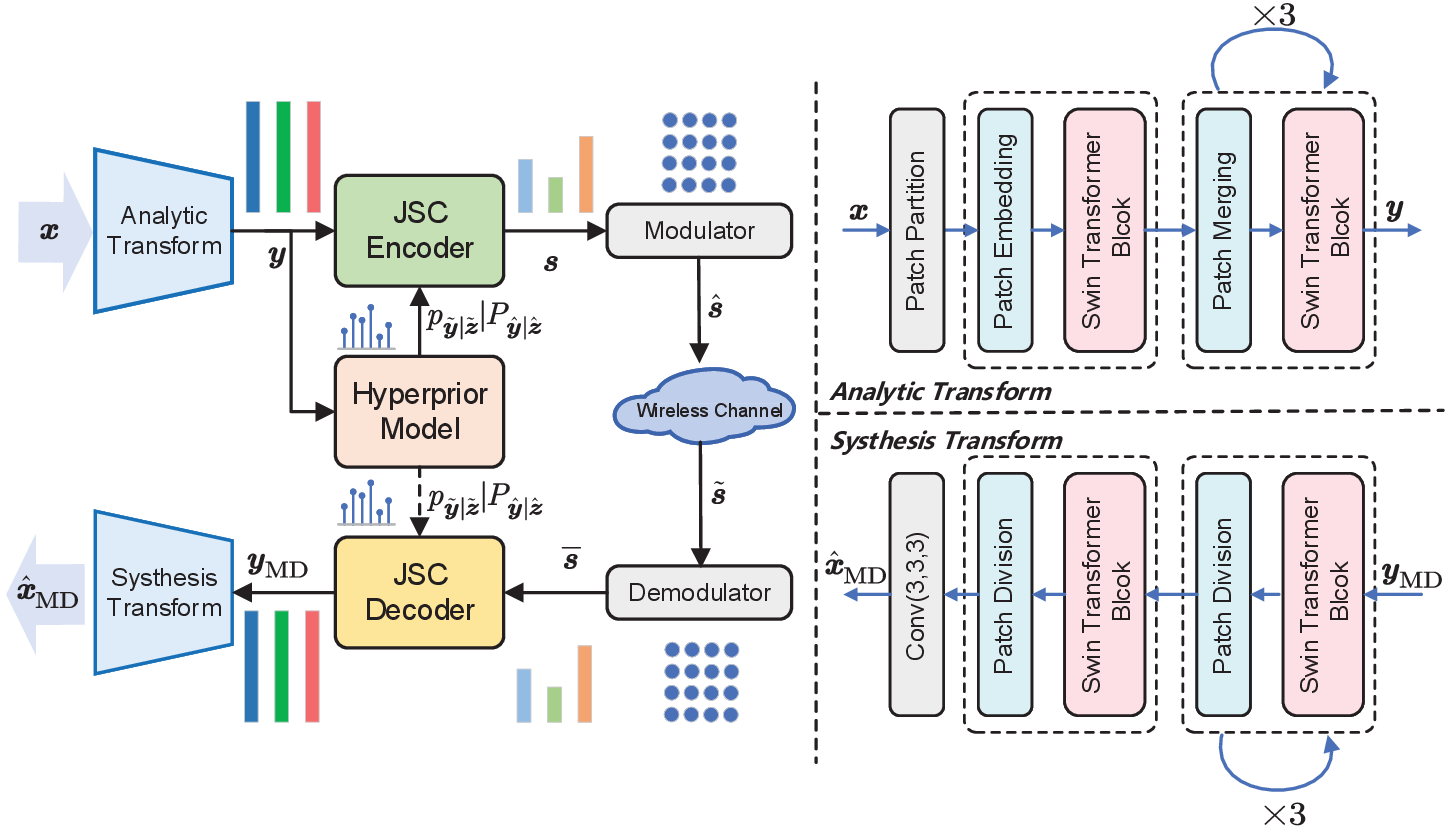}
		\caption{ The Overall architecture of MDJCM.} 
		\label{WholeModel}
	\end{centering}
\end{figure}
\begin{figure*}[t]
	\begin{centering}
		\includegraphics[width=0.95 \textwidth]{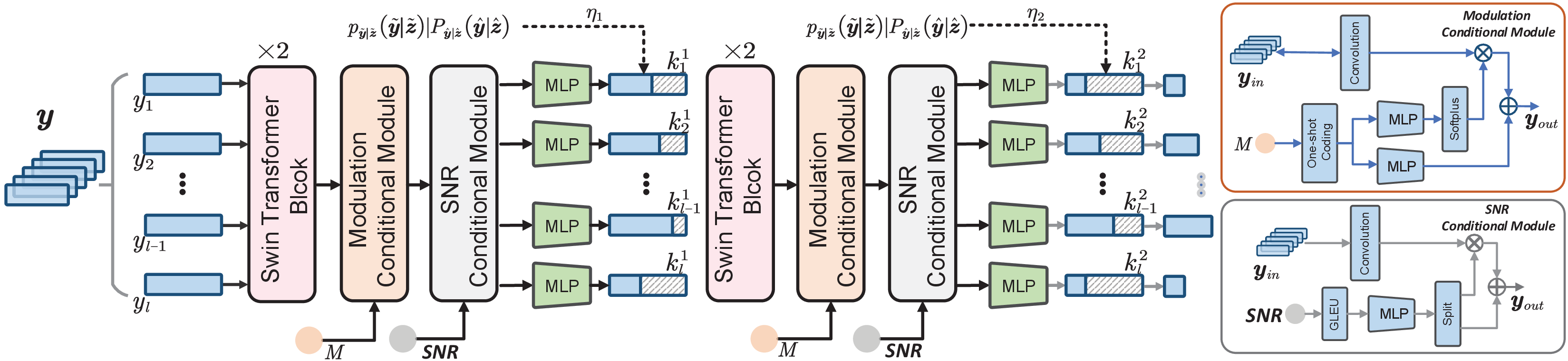}
		\par \end{centering}
	\caption{The illustration of the proposed SCM and MCM, as well as the hierarchical dimension reduction strategy. }
	\label{ModelArchitect}
\end{figure*}

\section{Model Architecture Design}\label{SEC4}
	In this section, we depict the design of MDJCM, including the details of the conditional network and hierarchical dimension reduction design. Finally, we provide a detailed introduction to the training process.

\subsection{Overall Model Architecture}

The overall architecture of our proposed MDJCM is shown in Fig. \ref{WholeModel}. As stated in the previous section, it comprises the analytic transform $g_a$, synthesis transform $g_s$, JSC encoder $f_e$, JSC decoder $f_d$, and the hyperprior model. The analytic transform and synthesis transform are mainly implemented with multiple Swin Transformer blocks, as depicted in Fig. \ref{WholeModel}. Besides, both the JSC encoder and decoder are designed with Swin Transformer blocks, which map the intermediate features to channel symbol vectors of different length. For the hyperprior model, we employ convolutional layer with a ReLU activation function.  

\subsection{Hierarchical Dimension Reduction}
In this work, we manage to control the number of transmission symbols for each image according to the estimated entropy. This is actually achieved by generating a mask vector $\bm{m}$ to mask the channel input vector, with certain elements being zero. Then, the encoded feature $\bm{s}$ output by the JSC encoder is masked by $\bm{m}$, given as $\bm{s} = \bm{s} \odot \bm{m}$, where $\odot$ denotes element-wise dot product. As a result, the transmission rate is controlled by reducing the output dimension. 
We find this one-time dimension reduction operation may lose effectiveness, since the model may suffer from heavy information loss in this sudden mask operation, which is caused by masking some important features by mistake. To address this issue, we propose a hierarchical dimension reduction mechanism, as shown in Fig. \ref{ModelArchitect}. Specifically, we employ multiple rate reduction operations, aiming to provide a gradual dimension reduction.  This approach effectively breaks down a complex selection problem into several simpler ones. By iteratively dropping a small subset of features, the model can dynamically adjust subsequent selections based on the remaining unmasked features. This adaptive process enables the model to tolerate errors more effectively compared to a single selection performed at the end of the process.

Specifically, the input of the JSC encoder $\bm{y}$ can be denoted as a vector sequence $(y_1, y_2,...,y_l)$, where each $y_i$ is a feature vector. The rate is controlled by masking portion of each vector, with the mask $\bm{m}$. To determine the mask ratio for each $y_i$, the entropy model is adopted. In particular, for each $y_i$, the length for the first time is set to $k^1_{i}=Q\left(-\eta_1 \log P_{\bar{y}_i |\bm{\bar{z}}}\left(\bar{y}_i | \bm{\bar{z}}\right)\right)$, where $\eta_1$ is employed to control the length of the feature $y_i$. According to $k^1_{i}$, we are able to generate the corresponding mask for $y_i$, and mask the intermediate feature. We consider using two reduction operations, where we introduce a new hyperparameter $\eta_2$ to further control the final length of $y_i$. Hence, the final length of $y_i$ is given by $k^2_{i}=Q\left(-\eta_2 \log P_{\bar{y}_i |\bm{\bar{z}}}\left(\bar{y}_i | \bm{\bar{z}}\right)\right)$, where $\eta_1 > \eta_2$. For each $y_i$, we generate two mask vectors
\begin{equation}
	\mathbf{m}_i^j=[\underbrace{1,1, \ldots, 1}_{k_i^j}, 0, \ldots, 0],\; \text{for} \; j=1,2.
\end{equation}
The generated vectors are adopted to mask $y_i$, resulting in a vector where some elements are set to zero.

\subsection{Conditional Network Design and Turbo decoder}
	We make an attempt to design a multi-order constellation generator that can produce constellation symbols of different orders. To avoid training different generators for different modulation orders, we design a modulation conditional module (MCM). This module can take the modulation order as conditioning input, accompanied by the intermediate feature. We integrate this module into the JSC encoder $f_a$, helping to adjust the encoding strategy for different modulation orders. Additionally, we aim to devise a fixed model capable of accommodating different channel conditions. To this end, we further consider an SNR conditional module (SCM), which takes SNR as input and produces the feature vector depending on the channel SNR.

	\begin{algorithm}[!htbp] \small
		\caption{Training the MDJCM} 
		\label{Training}
		\SetAlgoLined
		\textbf{Input:} Training data $\bm{x}$, the numbers of training steps for three phases $N_1$, $N_2$, and $N_3$, the scaling hyperparameters $(\eta_1, \eta_2)$, the Lagrange multiplier $\lambda$, the learning rate $l_r$, and the power constraint $\mathcal{E}_s$. \\
		\textbf{Output:} Parameters $\left(\bm{\theta}_g^*,\bm{\theta}_h^*,\bm{\theta}_f^*, \bm{\phi}_f^*,\bm{\phi}_h^*,\bm{\phi}_g^* \right)$. \\
		\textbf{First Phase: Train learning-based source coder}\\
		Randomly initialize the parameters and fix the parameters of JSC encoder and decoder, $\left(\bm{\theta}_f, \bm{\phi}_f\right)$.  \\
		\For{$i\leftarrow 1$ \KwTo $N_1$}{
			Sample $\bm{x} \sim p_{\bm{x}}$.\\
			Calculate the rate-distortion loss function $L(\bm{x},\bm{\hat{x}},\lambda)=R+\lambda D(\bm{x}, \bm{\hat{x}})$ based on (\ref{loss_func}). \\
			Update the parameters $\left(\bm{\theta}_g^*,\bm{\theta}_h^*,\bm{\phi}_h^*,\bm{\phi}_g^* \right)$.\\
		} 
		------------------------------------------------------------ \\
		\textbf{Second Phase: Learn the model with process (\ref{proprocess}).}\\
		Load the parameters trained in the first phase. \\
		\For{$i\leftarrow 1$ \KwTo $N_2$}{
			Sample $\bm{x} \sim p_{\bm{x}}$.  \label{step1}\\
			Sample SNR uniformly from the required range.   \\
			Randomly choose a modulation order $M$ from $\mathcal{M}$.\\
			Calculate $d$ according to the modulation order $M$ and power constraint $\mathcal{E}_s$. \label{step2}  \\
			Generate the reconstructed signal $\bm{\hat{x}}_{\text{MD}}$ through process (\ref{proprocess}). \\
			Calculate the loss function $L(\bm{x},\bm{\hat{x}},\lambda, M, \text{SNR})=R+\lambda [D(\bm{x}, \bm{\hat{x}})+D(\bm{x}, \bm{\hat{x}}_{\text{MD}})]$ based on (\ref{loss_func2}). \\
			Update the parameters $\left(\bm{\theta}_g,\bm{\theta}_h,\bm{\theta}_f, \bm{\phi}_f,\bm{\phi}_h,\bm{\phi}_g \right)$.} 
		
		------------------------------------------------------------ \\
		\textbf{Third Phase: Finetune the whole model with STE.}\\
		Load the parameters trained in the second phase. \\
		Fix the parameters of the transmitter $\left(\bm{\theta}_g,\bm{\theta}_h,\bm{\theta}_f \right)$.  \\
		\For{$i\leftarrow 1$ \KwTo $N_3$}{
			Repeat step \ref{step1} to step \ref{step2}.\\
			Generate the $\bm{\hat{x}}_{\text{MD}}$ through process (\ref{process}). \\
			Calculate the loss function $L(\bm{x},\bm{\hat{x}},\lambda, M, \text{SNR})=R+\lambda [D(\bm{x}, \bm{\hat{x}})+D(\bm{x}, \bm{\hat{x}}_{\text{MD}})]$ based on (\ref{loss_func2}). \\
			Calculate the gradient through STE. \\
			Update the parameters $\left(\bm{\phi}_f,\bm{\phi}_h,\bm{\phi}_g \right)$.}
	\end{algorithm}
	To implement MCM and SCM, we develop the architecture shown in Fig. \ref{ModelArchitect}. The main difference between MCM  and SCM lies in the processing of conditioning input, i.e., MCM employs a one-shot encoding while SCM uses a multi-layer perceptron (MLP). The developed conditional module mainly consists of a convolutional layer and MLP. The softplus operation is given by 
	\begin{equation}
		\textrm{softplus}(x) = \textrm{log}(1+e^x).
	\end{equation}
	By incorporating the modulation order and channel condition into the encoding process, the transmitter is able to generate the required signal accordingly. We integrate SCM and MCM into the JSC encoder and decoder, indicating the JSC encoder/decoder to adjust the encoding/decoding strategy dynamically based on the modulation order and channel condition.
	
	At the receiver, we employ the turbo decoder for signal denoising, which has proven to be effective in channel decoding \cite{Yihan_NIPS2019}. It consists of multiple decoder layers, each is comprised of a conv1d layer and a linear layer.
	
		\begin{figure}[t]
		\begin{centering}
			\includegraphics[width=0.26 \textwidth]{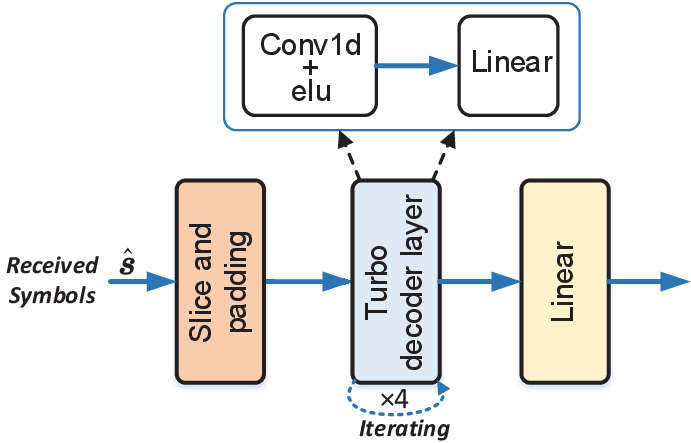}
			\caption{The architecture of the turbo decoder.} 
			\label{turbodecoder}
		\end{centering}
	\end{figure}
	
\subsection{Training Algorithm}
	The training process of MDJCM is concluded in Algorithm \ref{Training}.  Noting that MDJCM involves three main phases to optimize, i.e., the learning-based source coder, the JSC encoder/decoder, and the modulation modules, it is hard to train the model directly. To this end, we decompose the training process into different phases.
	At first, we begin with optimizing the learning-based coder, and thus only the parameters of the learning-based coder $\left(\bm{\theta}_g,\bm{\theta}_h,\bm{\phi}_h,\bm{\phi}_g \right)$ are trained, leaving the other modules fixed. This phase will give rise to a powerful learning-based coder, especially an accurate entropy model to indicate the rate determination for the JSC encoder.  In the second phase, we initialize the MDJCM with the parameters trained in the first phase and then learn the whole model with the process (\ref{proprocess}), using loss function (\ref{loss_func2}). In this phase, we will be able to obtain a powerful transmitter with great representation ability, as we have stated in Section III-C. Finally, we further finetune the receiver part of MDJCM with STE to eliminate the mismatch between the training process and the testing process, where the gradients are passed from the receiver to the transmitter. 
	Note that the overall training process may contain a number of training epochs.
	As MDJCM is expected to support various channel SNRs and modulation orders, during each training epoch, we randomly generate different SNRs and modulation order values. These values are then inputted into MDJCM, alongside sampled input source samples. In this way, MDJCM is able to generate corresponding channel symbols to deal with different channel conditions.
	
	\begin{figure}[!htbp]
		\begin{centering}
			\subfloat[]{\label{ov1}\includegraphics[width=2.8cm]{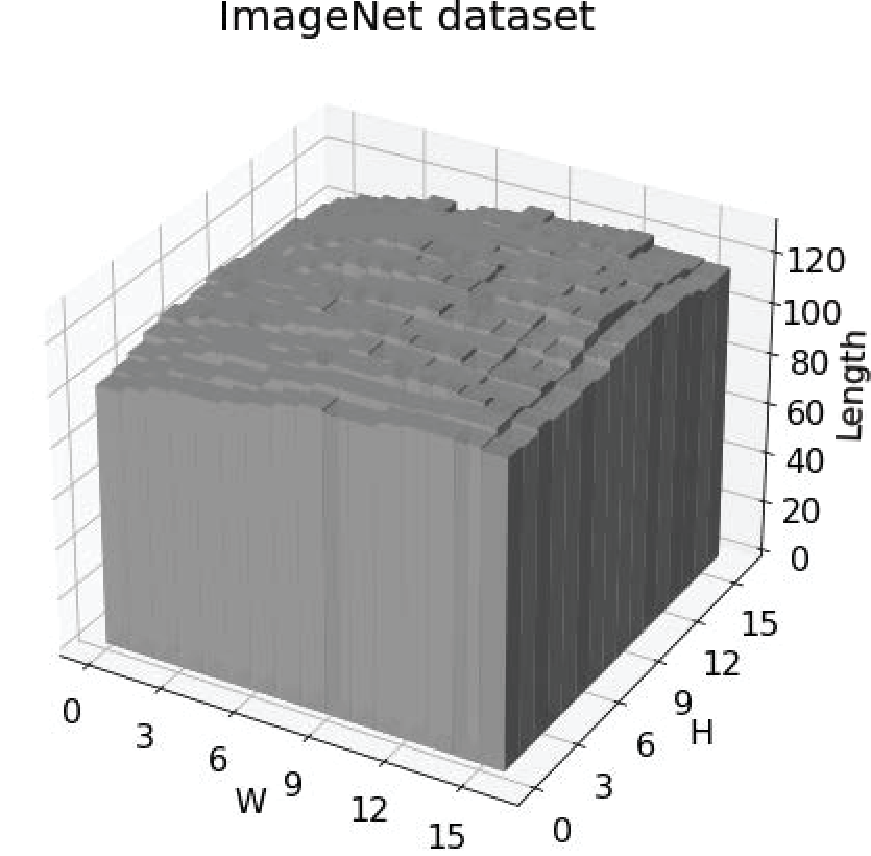}}  
			\subfloat[]{\label{ov2}\includegraphics[width=2.8cm]{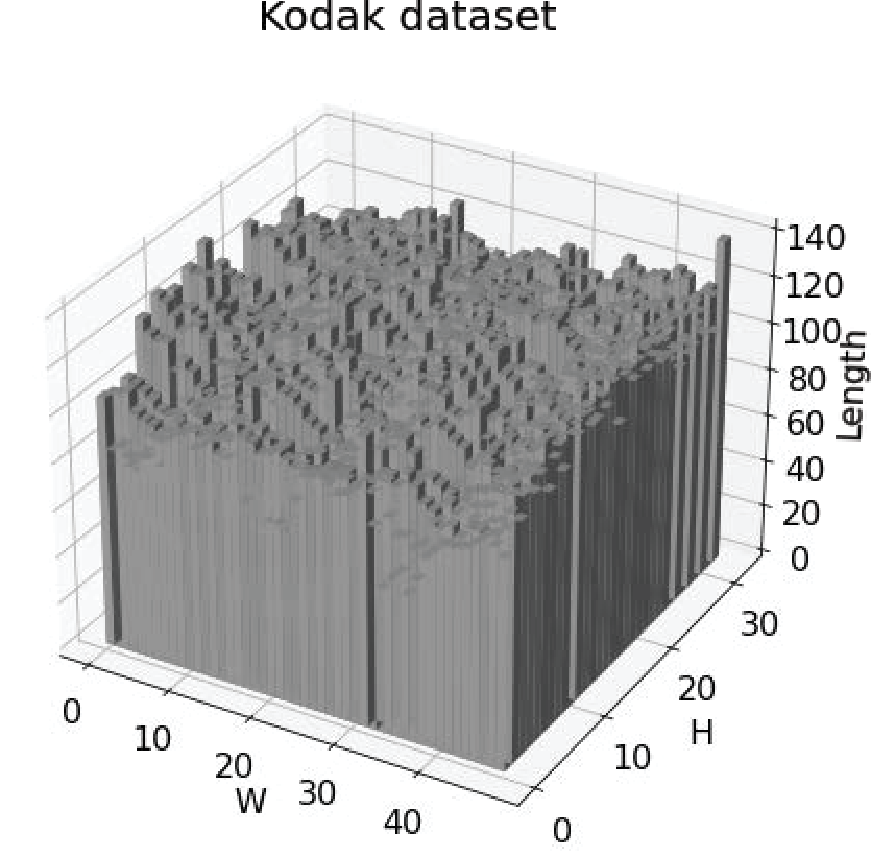}}
			\subfloat[]{\label{ov3}\includegraphics[width=2.8cm]{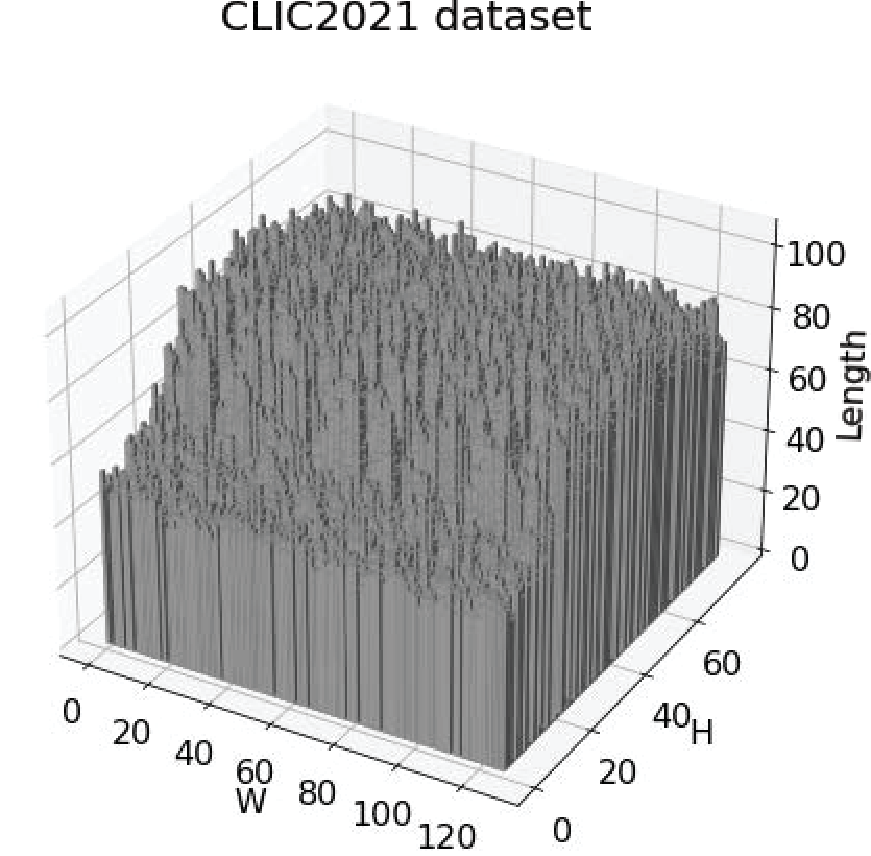}}
			\caption{The length (symbol count) distribution versus different datasets.} 
			\label{overhead_illu}
		\end{centering}
	\end{figure}
	Furthermore, we examine the resulted symbol length, $k^2_i$ for $i=1,2,...,l$, of the model trained with the imageNet dataset, as shown in Fig. \ref{overhead_illu}. We can observe a significant difference between the symbol length on imageNet and that on the CLIC2021 dataset, this training bias leads to a sub-optimal performance. Identifying the imageNet is stored with a lossy format, JPEG, we address this issue by incorporating the lossless DIV2K dataset as the training data, as what we will clarify in the Section V.

\begin{figure*}[t]
	\begin{centering}
		\subfloat[]{\label{overhead1}\includegraphics[width=4.5cm]{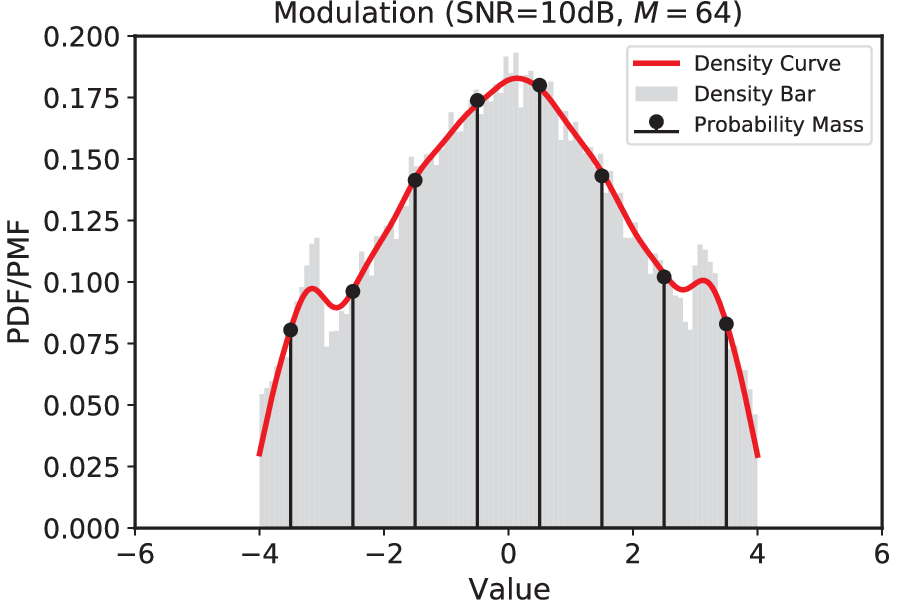}} 
		\subfloat[]{\label{overhead1}\includegraphics[width=4.5cm]{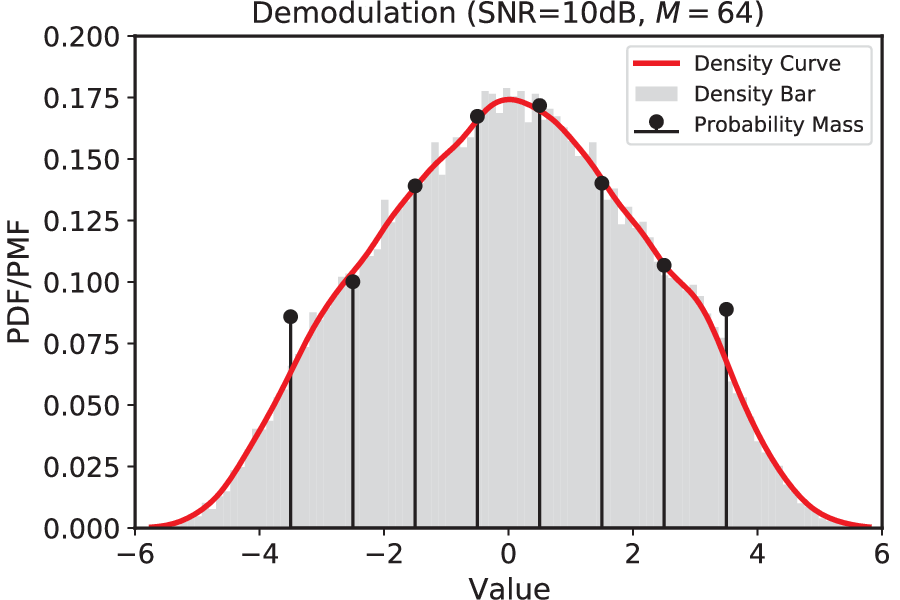}}
		\subfloat[]{\label{overhead1}\includegraphics[width=4.5cm]{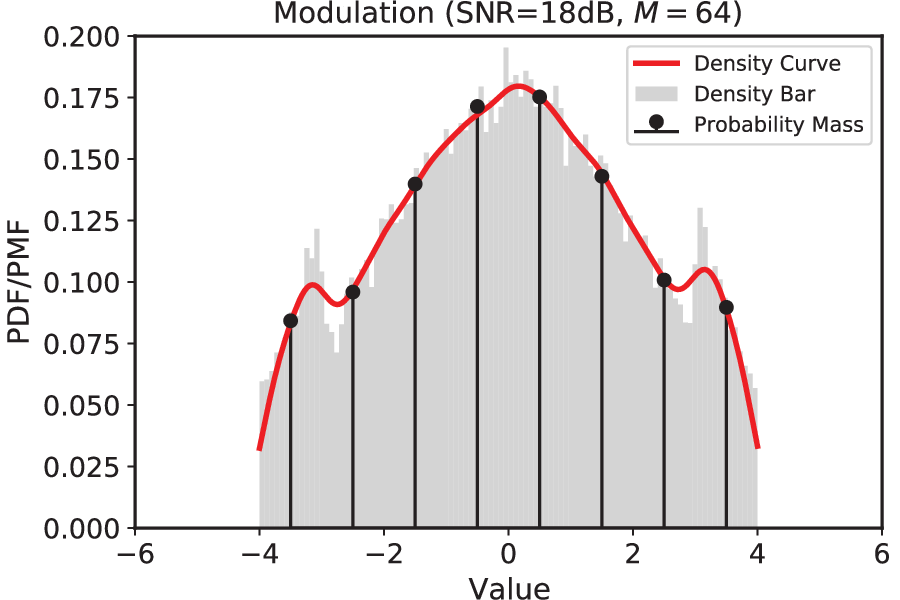}} 
		\subfloat[]{\label{overhead1}\includegraphics[width=4.5cm]{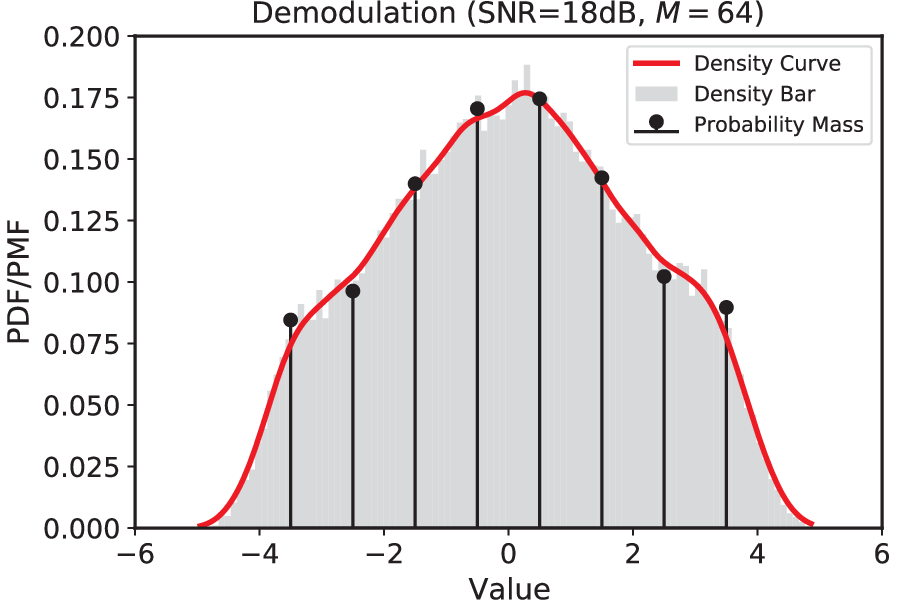}}\
		\subfloat[]{\label{overhead1}\includegraphics[width=4.5cm]{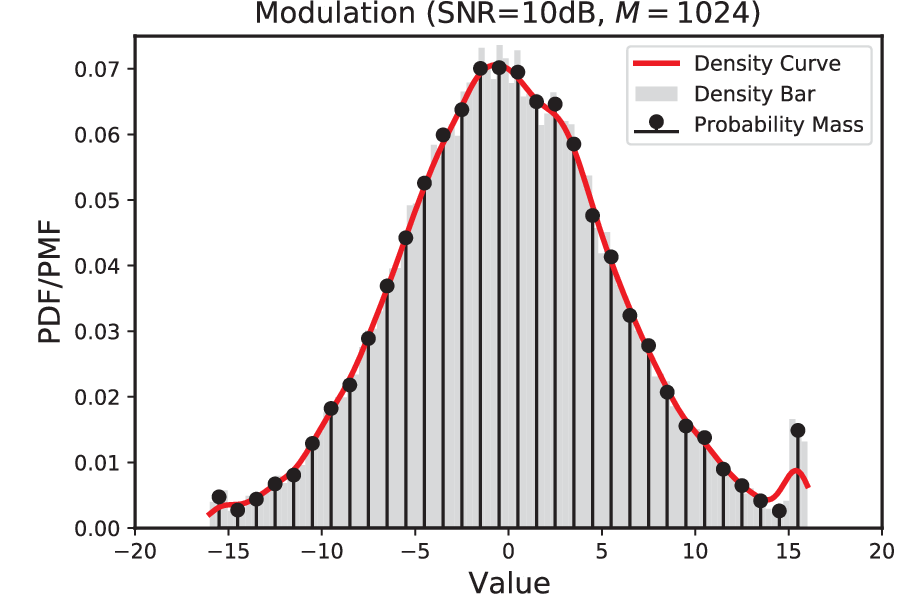}} 
		\subfloat[]{\label{overhead1}\includegraphics[width=4.5cm]{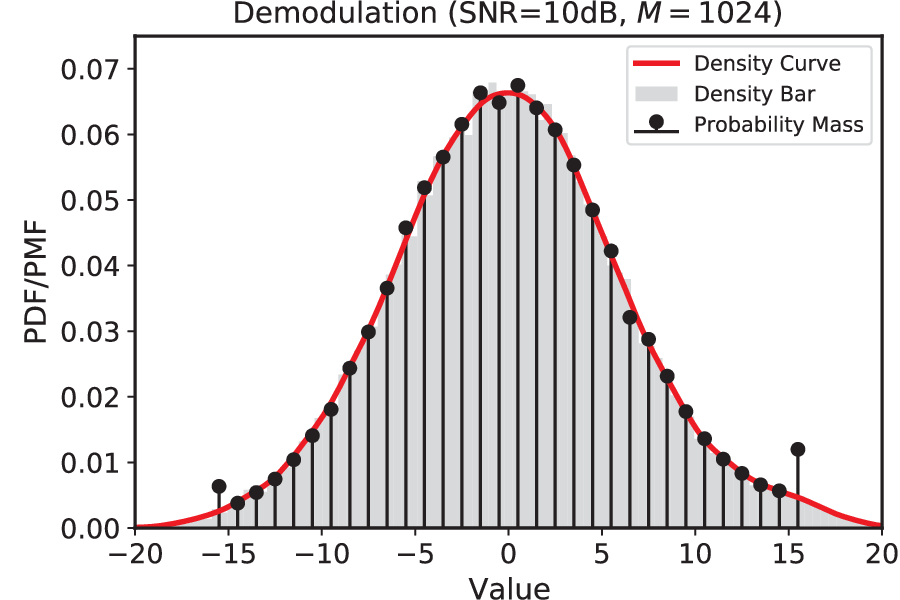}}
		\subfloat[]{\label{overhead1}\includegraphics[width=4.5cm]{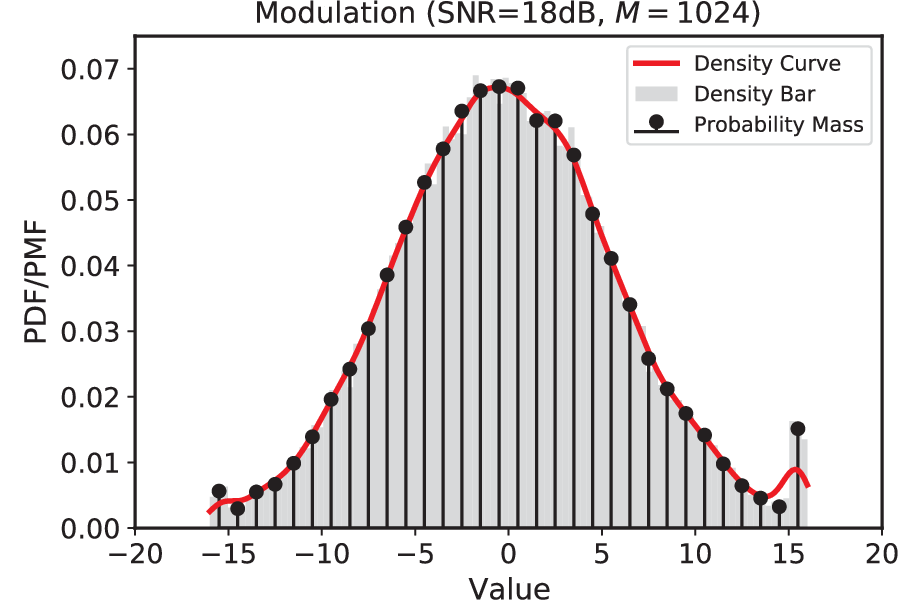}} 
		\subfloat[]{\label{overhead1}\includegraphics[width=4.5cm]{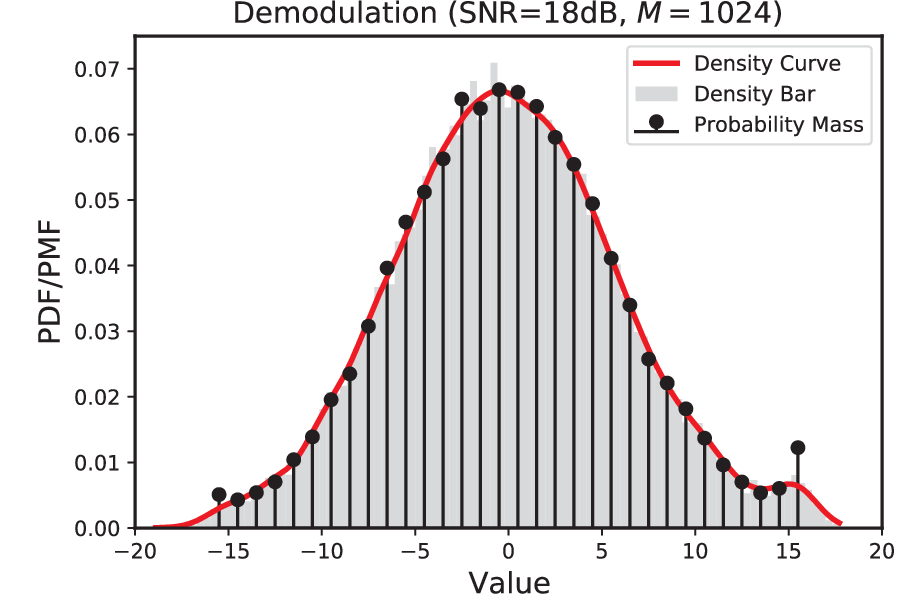}}
		\caption{The distribution comparisons after modulation and demodulation.} 
		\label{Density}
	\end{centering}
\end{figure*}

\section{Simulation Results}\label{SEC5}
\subsection{Simulation Settings}
	\subsubsection{Training Details} 
	We implement benchmarks and our proposed methods using PyTorch  \footnote{Our code and model weights are available in  \href{https://github.com/zhang-guangyi/Semantic-MDJCM.git}{https://github.com/zhang-guangyi/Semantic-MDJCM.git}.}. For training, we randomly select $50,000$ images with a resolution larger than $256\times 256$ from the ImageNet validation set and then randomly crop them to the size of $256 \times 256$ during training. We utilize the Adam optimizer with a batch size of $8$. The initial learning rate is set to $1\times 10^{-4}$. Notably, since ImageNet is in JPEG format, which represents a lossy version of the original images, we additionally employ the DIV2K dataset as the training dataset. The learning rate decays to $1\times 10^{-5}$ when training the model with DIV2K. The model is trained with a modulation set $\mathcal{M}=[4,16,64,256,1024]$, $\eta_1=0.4$, and $\eta_2=0.2$, we train the model at an SNR of $10$ dB. In addition, when we present the performance versus SNR, the SNR is uniformly drawn from the range $[0,13]$ dB. Moreover, we investigate the performance of the proposed methods in AWGN channels.

	\subsubsection{Evaluation}
	
	We evaluate our methods on various datasets with different resolutions.  To ensure fairness, we first include the CIFAR10 dataset. Additionally, we consider the Kodak dataset, which has an image size of $768\times 512$, as well as the CLIC2022 and CLIC2021 datasets,  both of which have a resolution of $2$k. These datasets encompass diverse resolutions and a wide range of real-world content, serving as common benchmarks for various image compression tasks.  To evaluate the performance of the proposed methods, we choose the peak noise-to-signal ratio (PSNR)    as the main metric, which measures the distortion between the transmitted image and the received image.  
	
	\subsubsection{Benchmarks}
				\begin{figure*}[!htbp]
			\begin{centering}
				\includegraphics[width=0.9 \textwidth]{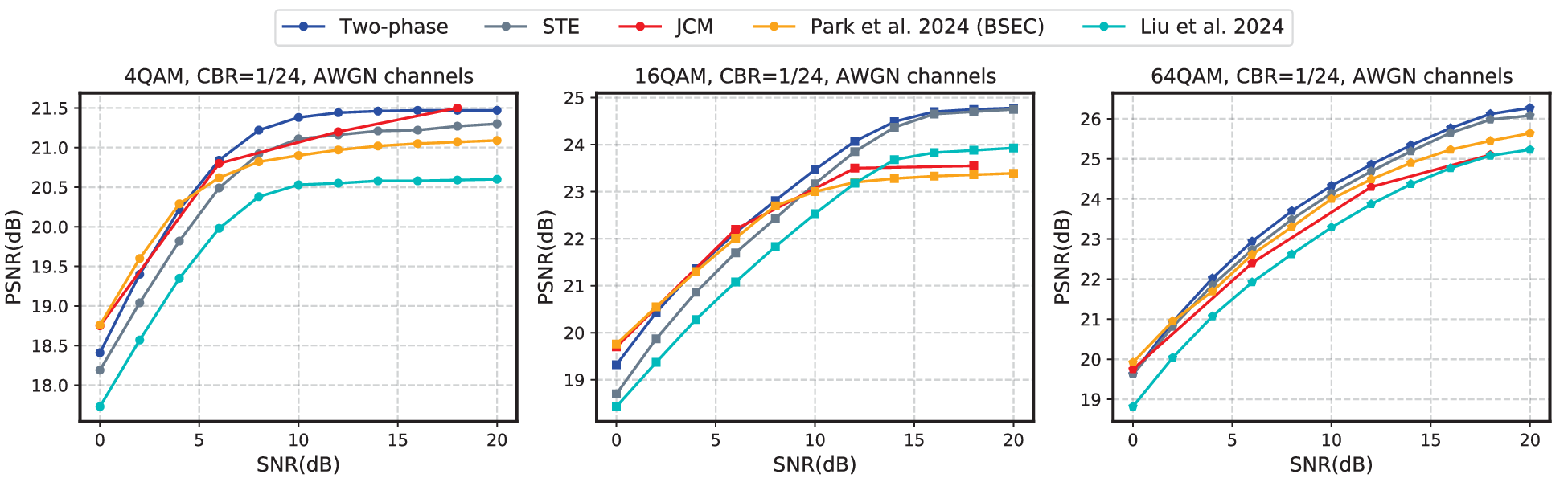}
				\caption{The PSNR performance versus SNR over CIFAR10 dataset.} 
				\label{PSNR_Overall}
			\end{centering}
		\end{figure*}
		\begin{figure}[!htbp]
		\begin{centering}
			\includegraphics[width=0.35 \textwidth]{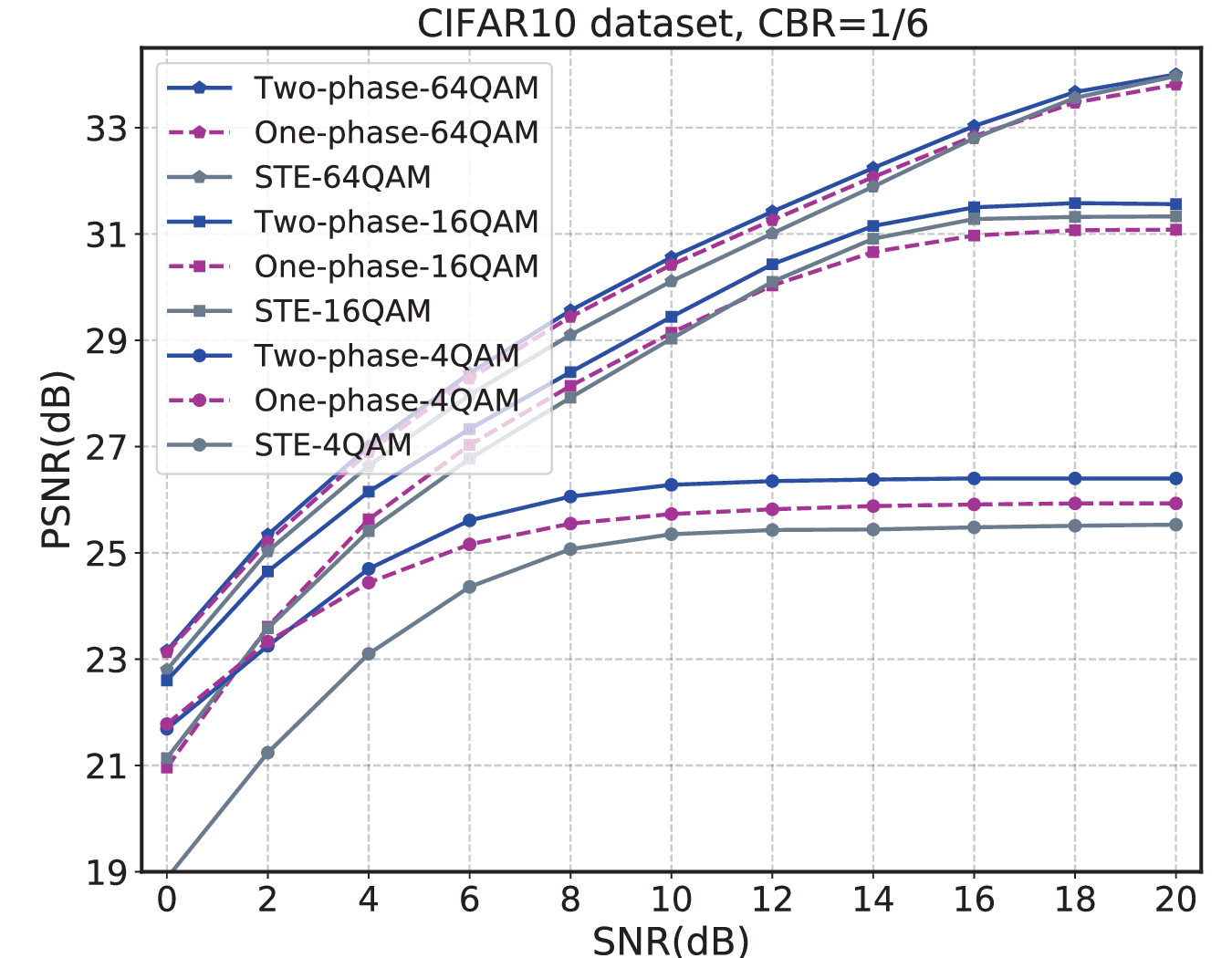}
			\caption{The relative performance gain of introducing the second phase to the training process by ablation study.} 
			\label{PSNR_CBR_STE}
		\end{centering}
	\end{figure}

	\begin{figure*}[t]
	\begin{centering}
		\subfloat[]{\label{overhead1}\includegraphics[width=5.5cm]{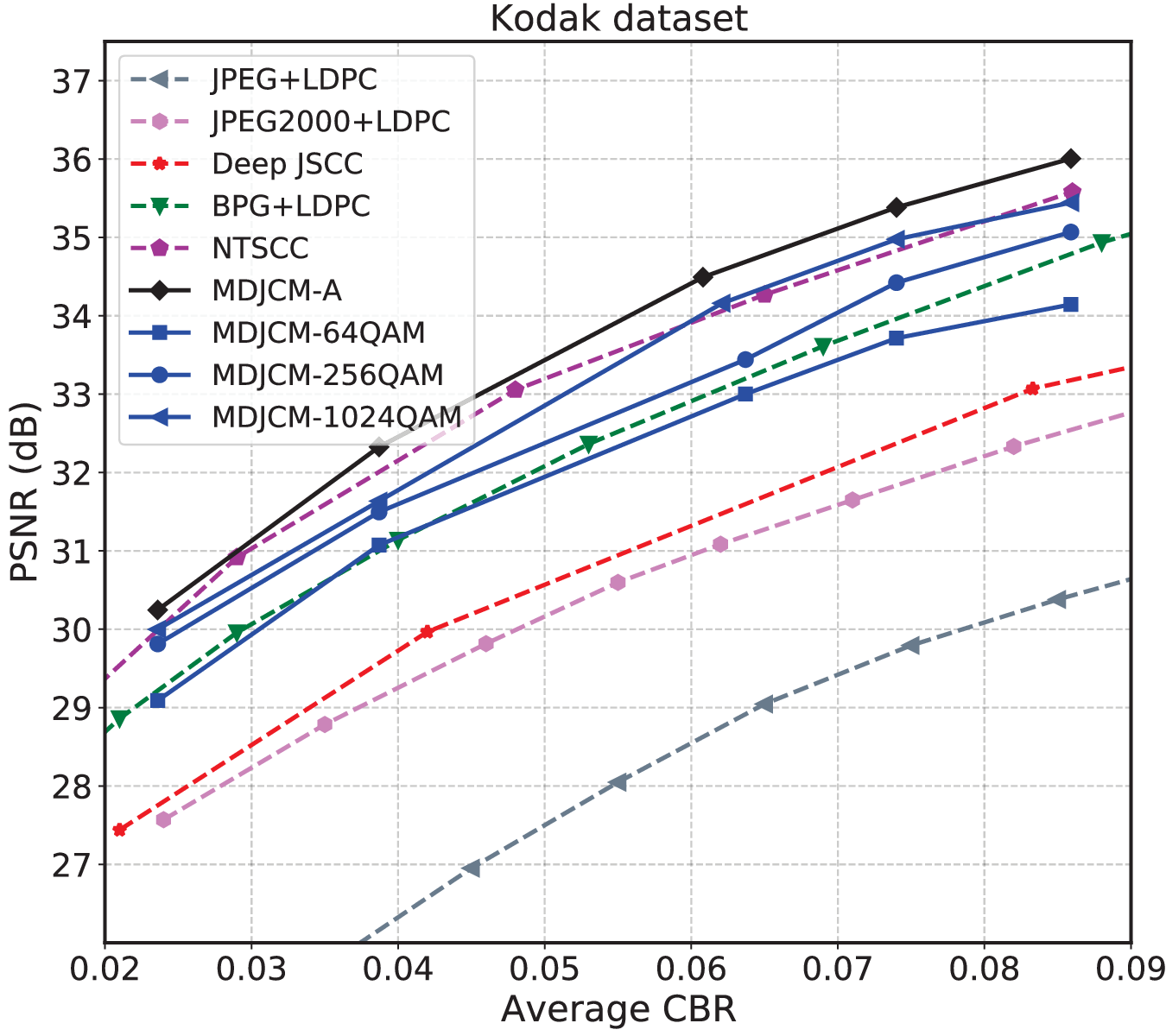}}
		\subfloat[]{\label{overhead1}\includegraphics[width=5.5cm]{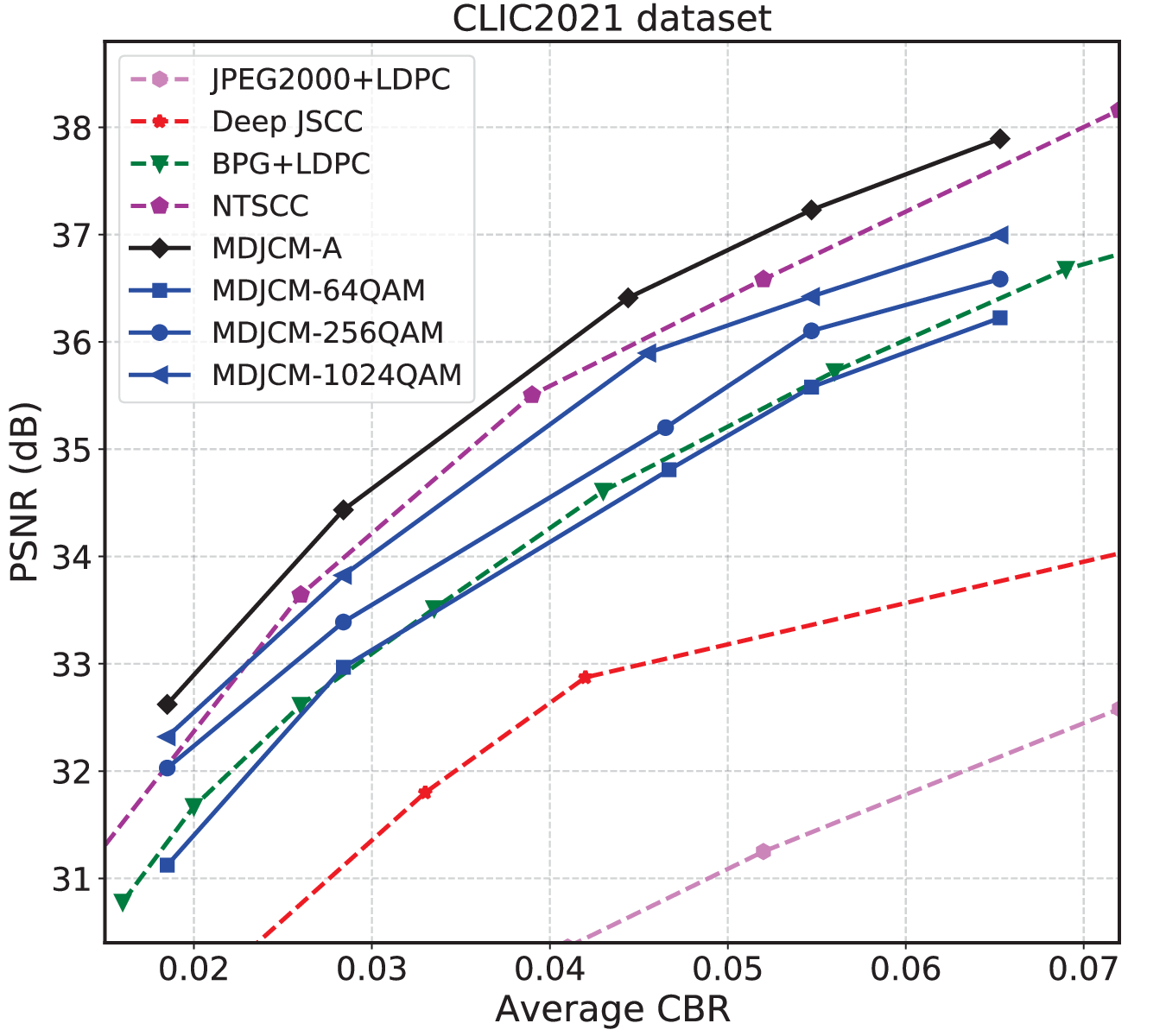}}
		\subfloat[]{\label{overhead2}\includegraphics[width=5.5cm]{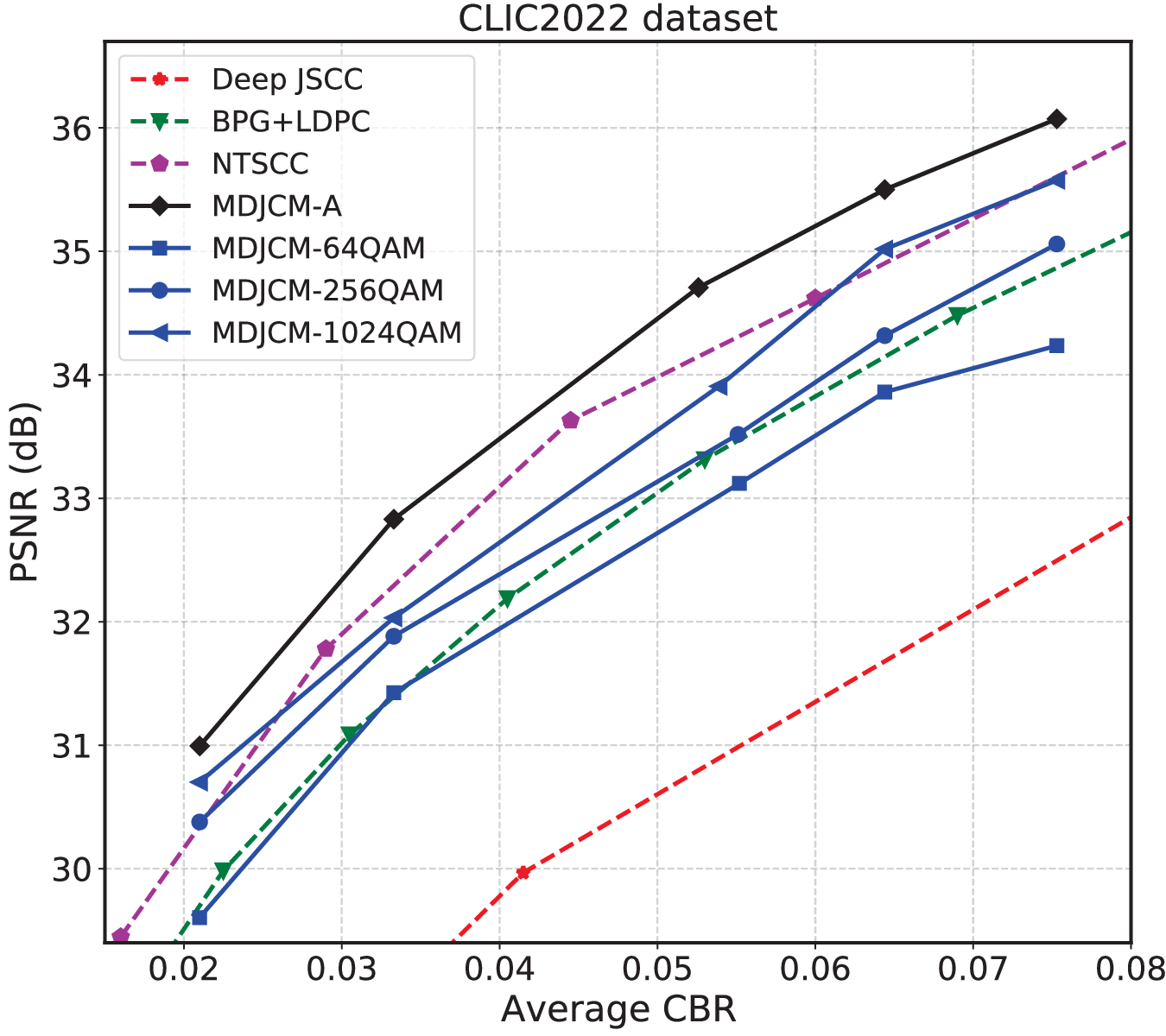}}
		\caption{The PSNR performance versus the average CBR over the AWGN channels at an SNR of $10$ dB. The considered datasets are (a) Kodak, (b) CLIC2021, and (c) CLIC2022, respectively.  } 
		\label{PSNR_CBR}
	\end{centering}
\end{figure*}
	For comparison, we incorporate several advanced mechanisms to demonstrate the effectiveness of the proposed methods. Firstly, we include the emerging DeepJSCC proposed in \cite{Eirina_TCCN2019}, where we adopt an improved model architecture to achieve superior performance compared to the original paper. 
	Additionally, we consider NTSCC \cite{Jincheng_JSAC2022} as a benchmark, showcasing the relative performance gain of the proposed methods. Furthermore, we include its variant, NTSCC+ \cite{Sixian_JSAC2023}, which achieves higher performance than NTSCC by introducing a context model to enhance coding efficiency. To the best of our knowledge, NTSCC+  represents the state-of-the-art deep learning-based analog JSCC model at present. 
	In addition to these analog JSCC systems, we introduce several digital methods. Specifically, we exploit the combination of the advanced image codec, BPG, with a practical channel codec, LDPC. The coding rate and modulation type are selected based on the SNR, ensuring approximately error-free transmission. For instance, when SNR $=10$ dB, the channel code rate is set to ${2}/{3}$ and modulation is selected as $16$QAM. 
	Moreover, the STE is also chosen as the digital baseline, where the transmitter directly outputs the constellation symbols, and the gradient is copied to the transmitter from the receiver to enable end-to-end training. Furthermore, we include the state-of-the-art joint coding-modulation (JCM) mechanism proposed in \cite{Yufei_Arxiv2023}, which realizes digital transmission based on a VAE. Moreover, in two recent works, \cite{Joohyuk_TCCN} and \cite{Chuanhong_OFDM}, the authors proposed methods for quantizing symbols into bits for transmission, which we also include for comparison. 
	
\subsection{Distribution Analysis}
	
	Notably, existing digital methods often concentrate on small-size image datasets such as CIFAR10. Therefore, we first incorporate CIFAR10 as the training dataset to reinforce the validity of our findings.
	As outlined in Section III-B, the probability density of $\bm{\bar{s}}$ in process (\ref{process}) is approximately proportional to the probability mass of $\bm{\bar{o}}$ using process (\ref{proprocess}). To validate this assertion, we conduct numerical Monte Carlo simulations. Specifically, we derive the PMF of $\bm{\bar{o}}$ through sample counting, while the PDF of $\bm{\bar{s}}$ is obtained by kernel density estimation. The simulation results are depicted in Fig. \ref{Density}. 
	In the first row, we present the results employing $64$QAM, under SNR conditions of $10$ dB and $18$ dB, respectively. The results of using $1024$QAM are shown in the second row. We provide the comparisons of the signal after modulation and demodulation. That is, we consider the comparison between $\bm{\hat{s}}$ and $\bm{\bar{s}}$, as well as the comparison between  $\bm{\bar{o}}$ and $\bm{\hat{o}}$. From the results, we can clearly see that the proposed process (\ref{proprocess}) can approximate the process (\ref{process}). This approximation is quite accurate at different SNRs, especially at the high SNR regime. Furthermore, we note a slight approximation error, primarily noticeable at the edge constellation symbols. However, this discrepancy remains minor, ensuring the continued effectiveness of (\ref{proprocess}) across various conditions. Notably, we observe that this approximation achieves greater accuracy with higher modulation orders, consistent with the rationale detailed in Section III-B. By introducing the two-phase training strategy, the impact of this error is significantly eliminated.

\subsection{Performance Analysis}
	
	In Fig. \ref{PSNR_Overall}, we present a comparison of the PSNR results obtained using different training methods across varying SNR levels under the AWGN channel. Specifically, we introduce the STE  as a benchmark to demonstrate the relative performance gain achieved by our proposed two-phase training strategies. Additionally, we include as a benchmark the advanced digital joint coding-modulation (JCM) proposed by \cite{Yufei_Arxiv2023}, which employs the VAE to output the probabilities of each constellation symbol and sample from the probabilities to generate the constellation symbols. Since there is no released code from \cite{Yufei_Arxiv2023}, we implement a similar architecture to that reported in \cite{Yufei_Arxiv2023} for different schemes. Additionally, we also incorporate two more recent methods, \cite{Joohyuk_TCCN} and \cite{Chuanhong_OFDM}, as benchmarks, which are labeled as ``Park et al. 2024 (BESC)" and ``Liu et al. 2024" in Fig. \ref{PSNR_Overall}, respectively.
	From the results, we find the proposed two-phase strategy significantly outperforms JCM and training with STE, where STE holds a higher performance than JCM at most modulation orders. We can also observe a performance saturation for these  schemes since transmitting with fixed modulation order introduces a bound for the amount of information. For instance, when SNR is larger than $8$ dB, almost no extra redundancy will be introduced, and the performance will not improve with SNR as already all of the transmitted symbols have been employed for source representation. Moreover, Park et al. 2024 (BESC) show competitive performance at a low SNR regime while decreased performance when SNR is high, especially when 4QAM is employed. It mainly stems from its mixed training strategy, where the model tends to maintain an excessive robustness against different noise levels, and thus fails to achieve high performance in the whole SNR range. Since the model in Liu et al. 2024 does not consider the modulation process in the training phase, it achieves the worse performance.
	
	From Fig. \ref{PSNR_CBR_STE}, it is evident that training with the substitution process (\ref{proprocess}) yields significantly higher PSNR performance compared to direct training with STE. The performance gain is much more significant when employing lower modulation orders and operating in low SNR regimes.  
	Specifically, training with the process (\ref{proprocess}) outperforms STE when SNR is below $12$ dB using $4$QAM. Through the utilization of the two-phase strategy, performance can be further enhanced, surpassing that achieved by STE comprehensively, even in high SNR scenarios.	
	Fig. \ref{PSNR_CBR_STE}(a) and (b) illustrate the effectiveness of the proposed methods across various CBR values, further confirming their generalizability.

	\begin{figure}[t]
	\begin{centering}
		\includegraphics[width=0.35 \textwidth]{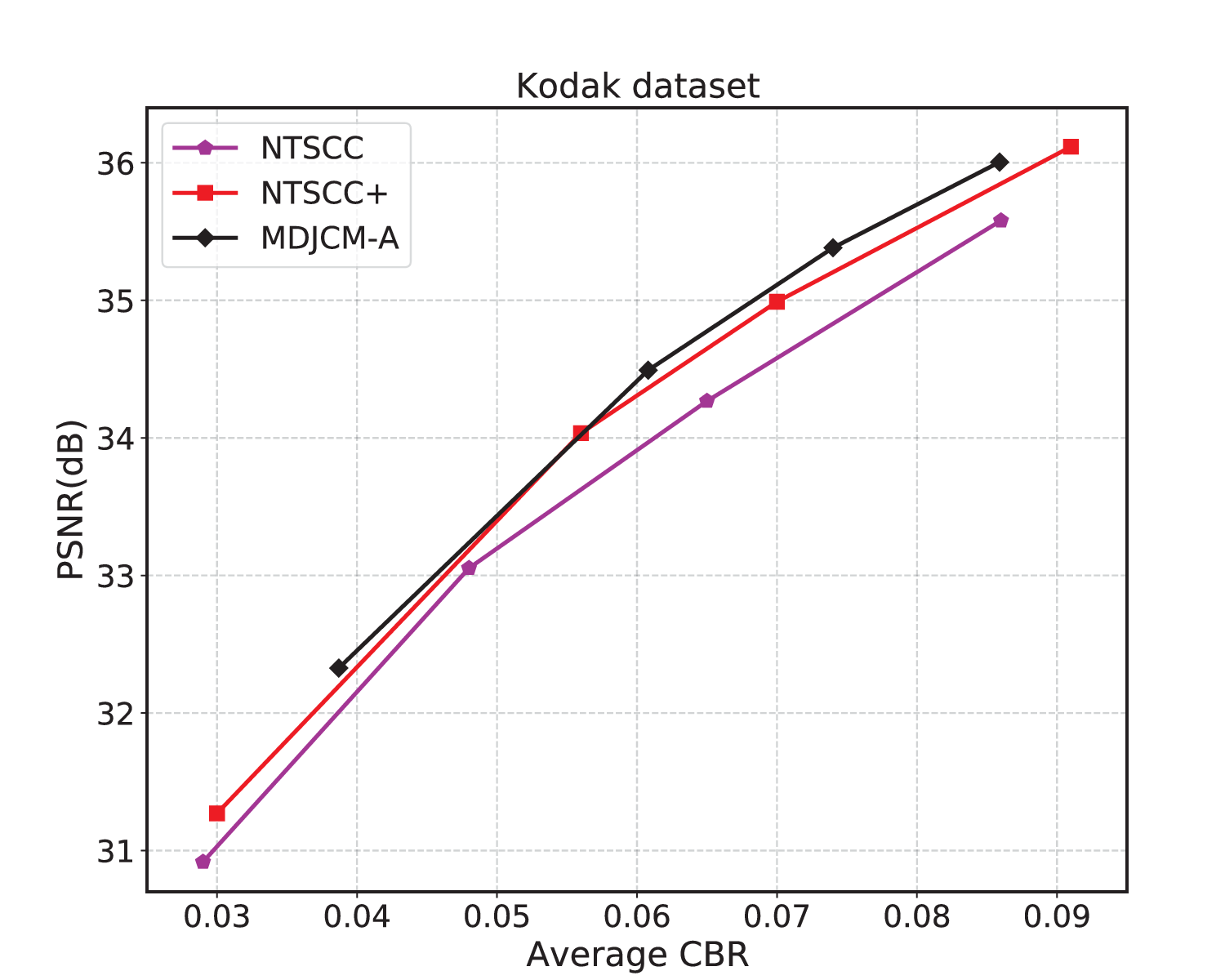}
		\par \end{centering}
	\caption{The PSNR performance versus the CBR over the AWGN channels, where all the schemes employ the analog transmission to show the relative performance gain.}
	\label{Hierarchical}
\end{figure}

		\begin{figure}[t]
		\begin{centering}
			\subfloat[]{\label{o}\includegraphics[width=4.5cm]{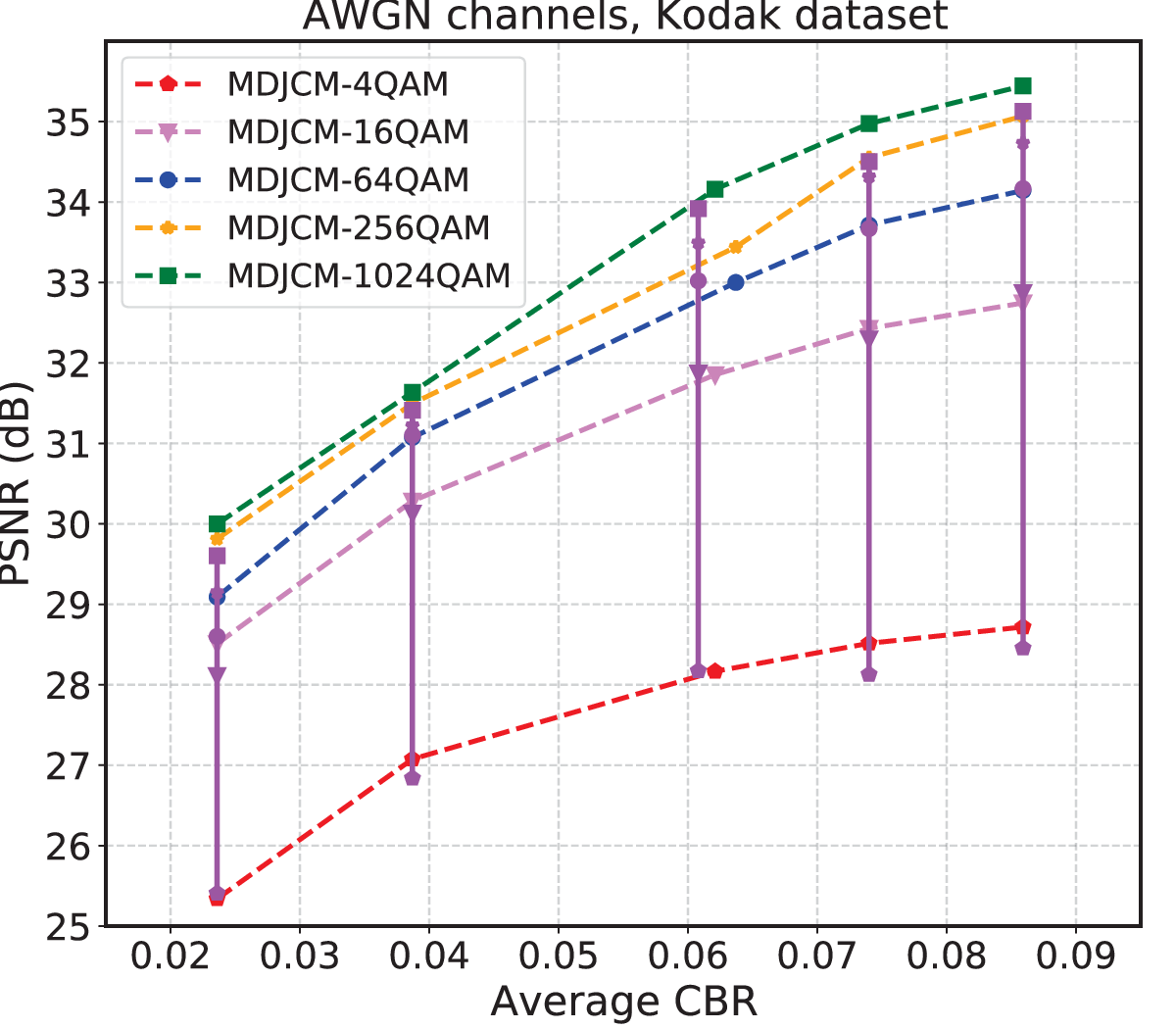}}
			\subfloat[]{\label{}\includegraphics[width=4.5cm]{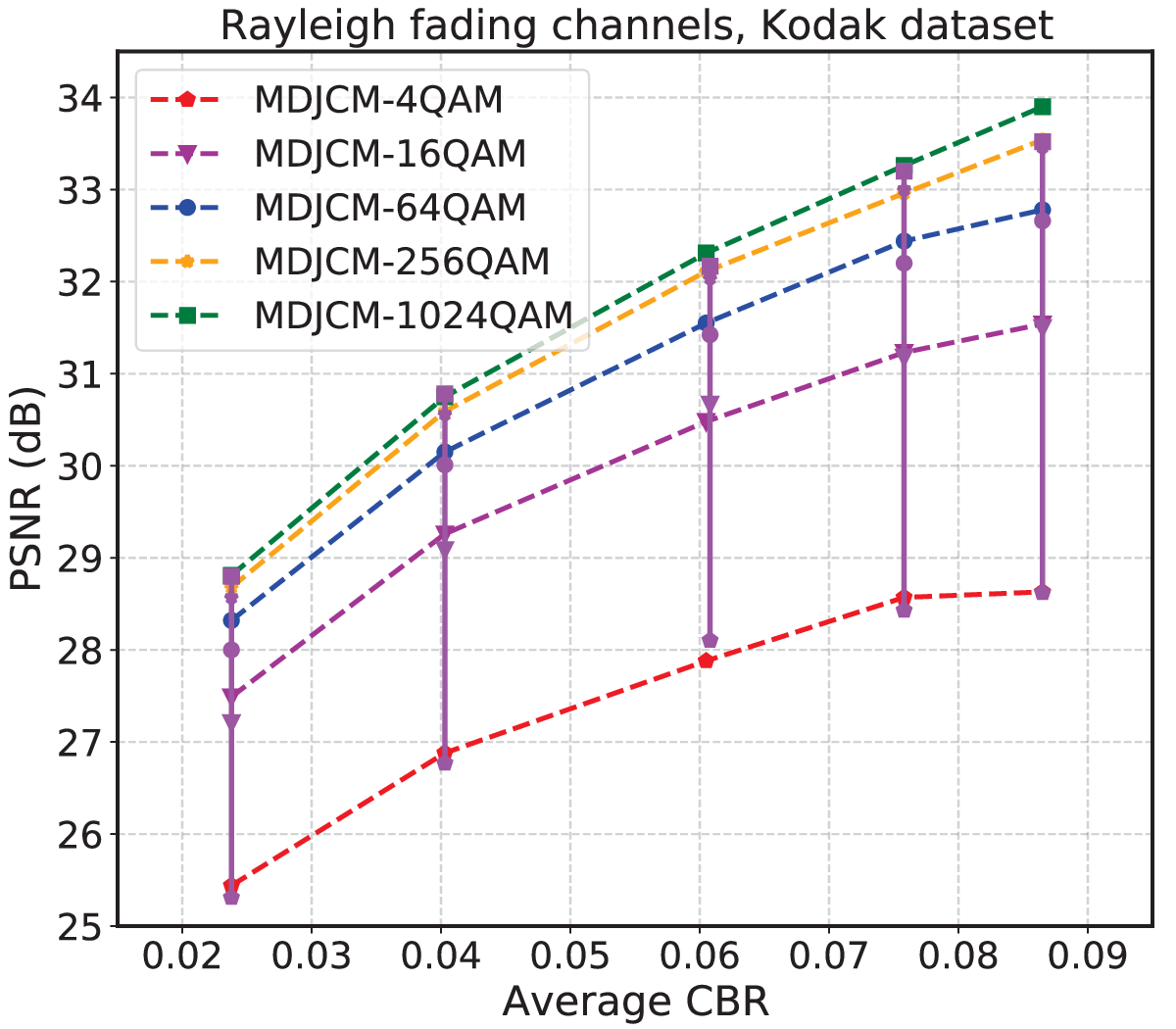}}
			\caption{PSNR performance versus CBR of training with different modulation orders. A comparison between the multi-order modulator/demodulator and a fixed-order modulator/demodulator. } 
			\label{PSNR_CBR_SCM}
		\end{centering}
	\end{figure}

	Fig. \ref{PSNR_CBR} shows the PSNR performance versus CBR over the AWGN channels at an SNR of $10$ dB. In this figure, "MDJCM-A" represents the utilization of analog transmission for the proposed MDJCM, serving as an upper bound. In addition, we consider $64$QAM, $256$QAM, and $1024$QAM in this figure, where the model is only trained with fixed modulation order. Our findings indicate the superiority of the proposed methods over DeepJSCC and traditional separation-based schemes like JPEG+LDPC, JPEG2000+LDPC, and BPG+LDPC. Notably, DeepJSCC experiences performance deterioration with increasing CBR due to its fixed-rate nature, unlike MDJCM and NTSCC, which leverage adaptive rate allocation for each image. Besides, the Swin Transformer employed in MDJCM is much more effective than the standard CNN in DeepJSCC. Moreover, we observe a greater performance gain at higher CBR levels compared to lower ones, as higher CBR allows for more flexible transmission rate allocation among different images. For the three high-resolution datasets, we can all observe the relative performance gain. Furthermore, in training with higher modulation order, the performance of MDJCM will approach that of MDJCM-A, since the modulation error decreases with the modulation order. This phenomenon also explains why the performance using $1024$QAM surpasses that of $64$QAM and $256$QAM.

	In Fig. \ref{Hierarchical}, we compare the analog transmission performance of the proposed MDJCM with NTSCC and its advanced variant, NTSCC+. The primary distinction between MDJCM-A and NTSCC lies in the hierarchical dimension reduction strategy introduced in MDJCM-A. NTSCC+ represents an enhanced version of NTSCC, incorporating a context model to improve coding efficiency, which was considered as state-of-the-art in wireless image transmission. Our observations from the figure reveal that the proposed methods significantly outperform NTSCC, underscoring the effectiveness of our approach. Moreover, MDJCM-A is even able to surpass NTSCC+, particularly in the high CBR region. This also reveals the necessity to design a powerful JSCC part in such a variable-rate scheme, while NTSCC+ mainly focuses on the learning-based source coder by employing a more advanced source coder.
	
	\begin{figure*}[t]
		\begin{centering}
			\subfloat[]{\label{overhead1}\includegraphics[width=6.6cm]{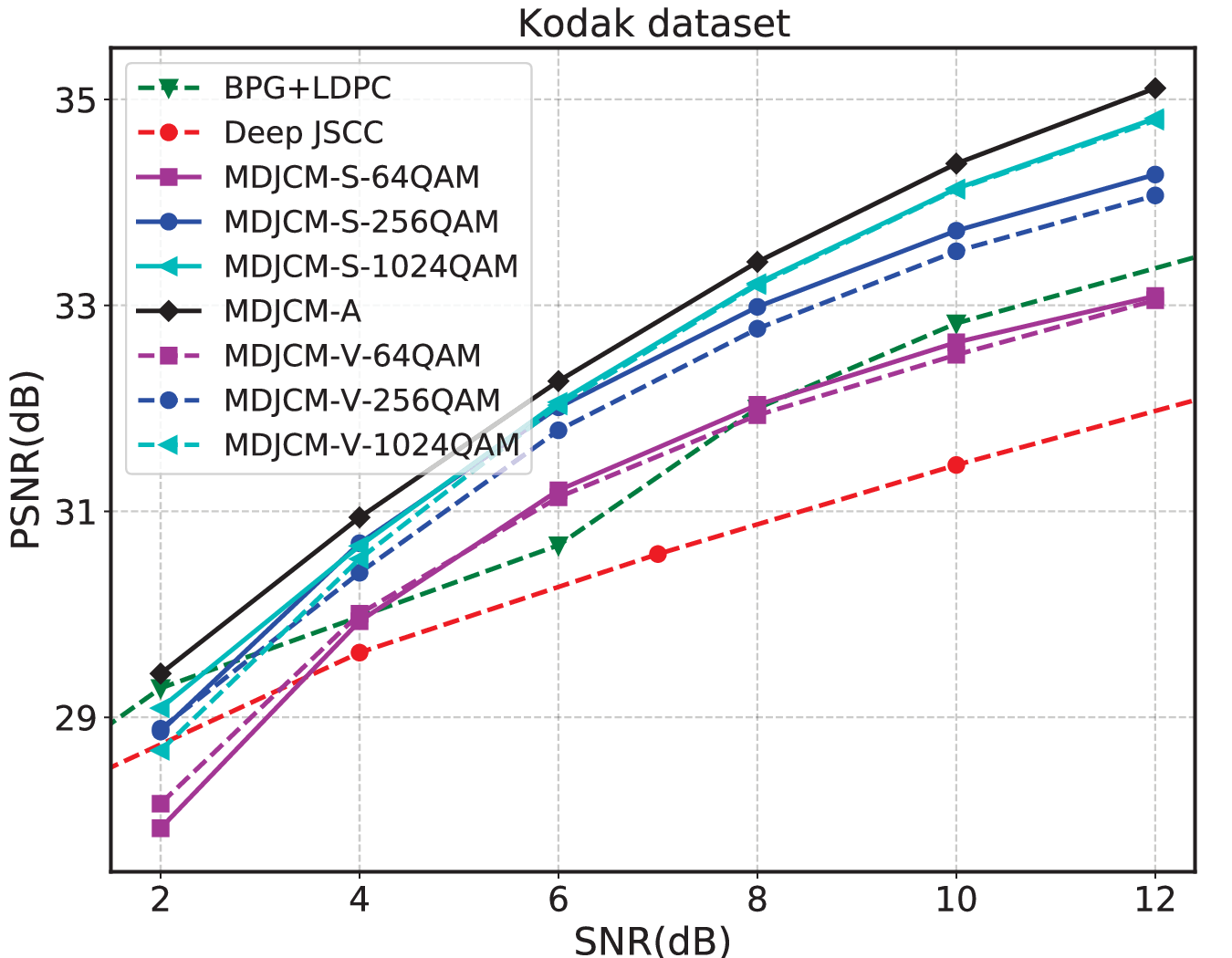}} \quad
			\subfloat[]{\label{overhead1}\includegraphics[width=6.6cm]{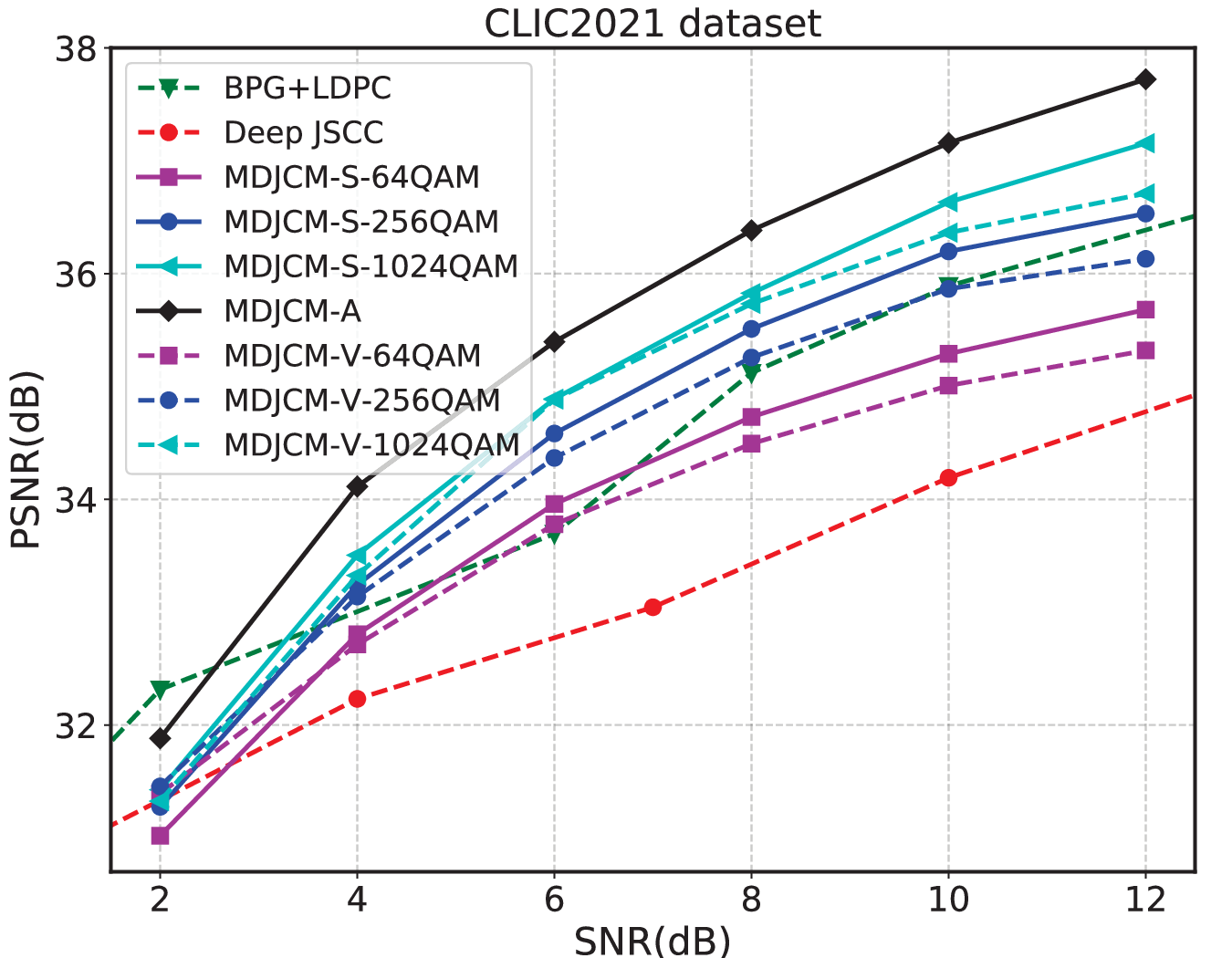}}
			\caption{The PSNR performance versus the SNR over different datasets. The average CBR is set to $0.0625$.} 
			\label{PSNR_SNR}
		\end{centering}
	\end{figure*}
		\begin{figure*}[t]
		\begin{centering}
			\includegraphics[width=0.98\textwidth]{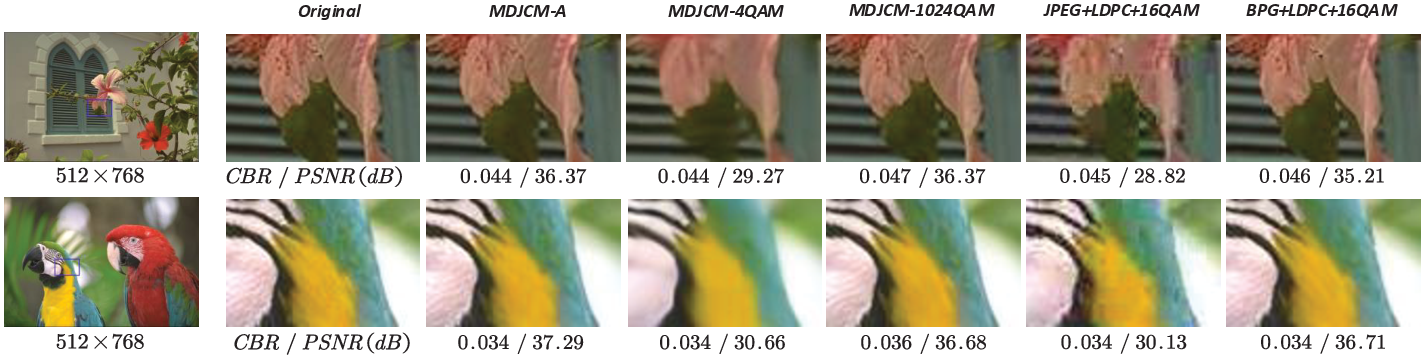}
			\par \end{centering}
		\caption{Visualization example of the reconstructed images. The metrics are [CBR/PSNR]. These reconstructed images are obtained by employing different schemes over the AWGN channel at the SNR of $10$ dB.}
		\label{VisionExample}
	\end{figure*}
	
	In Fig. \ref{PSNR_CBR_SCM} (a), we present the results of training the model with varying modulation orders $\mathcal{M}=[4,16,64,256,1024]$, where the performance is indicated by the black points. The other curves are obtained by training the MDJCM with a fixed modulation order. The dataset considered is Kodak, and the channel models include AWGN and Rayleigh fading. From the results, we can observe only a slight performance degradation when using a multi-order constellation generator, underscoring the efficacy of our proposed conditional network, where our proposed multi-order constellation generator can approach the specifically trained model. In addition, the performance degradation mainly exists in higher modulation order and higher CBR regions. For example, the performance of MDJCM shows comparable performance to the specifically-trained one, except for $1024$QAM and CBR greater than $0.07$. Moreover, the performance gap between different modulation orders widens with increasing CBR, highlighting the limitations of lower modulation orders in supporting higher CBR. Compared to fixed modulation schemes, our proposed approach offers greater compatibility with contemporary digital communication systems, as there is no need to train multiple models to support different requirements. 
	Additionally, we include results for Rayleigh fading channels in the simulation section, as shown in Fig. \ref{PSNR_CBR_SCM} (b). This figure illustrates the robustness of the proposed MDJCM across varying channel SNRs. We observe that our MDJCM performs exceptionally well in different SNR regimes on Rayleigh fading channels, highlighting its generalizability across diverse channel environments.

	Fig. \ref{PSNR_SNR} (a) and (b) depict the PSNR performance as SNR varies, utilizing the Kodak dataset and CLIC2021 dataset, respectively, while maintaining a constant CBR value of $0.0625$ for all transmission schemes. The figure includes the traditional separation scheme, BPG+LDPC, where modulation and channel code rate adapt according to the channel SNR. Next, we also consider the DeepJSCC, which is trained at a specific SNR and should be able to achieve its best at each SNR point. 
	For our proposed scheme, we train the model across a broad SNR range using either a fixed modulator, denoted as "MDJCM-S", or a multi-order modulator, denoted as "MDJCM-V". Our results demonstrate the capability of our proposed scheme to support applications over a wide channel SNR range without requiring multiple models trained at specific SNR conditions. Notably, MDJCM exhibits superior performance compared to DeepJSCC trained at specific SNR conditions.
	Moreover, traditional separation-based digital methods like BPG+LDPC often suffer from the cliff effect when SNR falls below the expected level by the channel code. In contrast, MDJCM provides gradual degradation as SNR decreases and even approaches analog system performance at higher SNR regions due to its flexibility in adjusting the tradeoff between channel and source coding.
	Comparing MDJCM-S with MDJCM-V, we observe only slight performance degradation despite MDJCM-V needing to handle multiple modulation orders, underscoring the effectiveness of our proposed methods.

	To further illustrate the performance of the proposed methods, Fig. \ref{VisionExample} presents examples of the reconstructed images using different transmission schemes. Compared with the benchmarks, the proposed methods provide better visual quality with lower CBR cost. In addition, applying higher modulation order can achieve better reconstruction quality. In addition, the proposed methods can help to keep more texture and details than the BPG+LDPC. This further demonstrates the potential of employing the proposed methods in wireless image transmission.
	
\section{Conclusion} \label{Conclusion}\label{SEC6}
	In this work, we have proposed a modulation-agnostic digital semantic communication system for wireless image transmission. Starting from the learning-based coder, we formulated a framework for digital systems capable of achieving variable-rate transmission. To overcome the challenges posed by non-differential modulation/demodulation processes, we introduce a novel substitution training strategy. This strategy treats them as limited and scaled quantization processes, with an explanation of their equivalence provided. To eliminate the mismatch between the training and testing phases, we introduce a two-phase training strategy. For the realization of a variable-order modulator, we propose a conditional network that instructs the transmitter to generate corresponding constellation symbols.  Additionally, we implement a hierarchical dimension-reduction strategy to further enhance performance. In essence, our paper presents a promising and effective method for training a digital semantic communication system in an end-to-end fashion, exhibiting superior performance compared to existing solutions.

\begin{appendices}

\section{Derivation of (\ref{denisty})}
	
	According to the probability convolution formula, we have
		\begin{equation}
		p_{\bar{o}_{i}^{In}}(u) = p_{s_i^{In}}(u)*\frac{1}{2d}\text{rect}(\frac{u}{2d}     )*p_{e_i^{In}}(u)*\frac{1}{2d}\text{rect}(\frac{u}{2d}     ).
	\end{equation}
	Since
	\begin{equation}
		p_{s_{i}^{In}}(u)*\frac{1}{2d} \text{rect}(\frac{u}{2d}    ) = \int^{u+d}_{u-d}p_{s_{i}^{In}}  ({\mu}) \frac{1}{2d} d\mu.
	\end{equation}
	and
	\begin{equation}
	p_{e_i^{In}}(u)*\frac{1}{2d}\text{rect}(\frac{u}{2d}    ) = \int^{u+d}_{u-d}p_{e_i^{In}}(\tau) \frac{1}{2d}  d\tau,
	\end{equation}
	we have
	\begin{equation}
	\begin{aligned}
		& p_{\bar{o}_{i}^{In}}(u)=\\
		& \int_{-\infty}^{\infty} \int^{\nu+d}_{\nu-d}p_{s_{i}^{In}}(\mu)  \frac{1}{2d}  d\mu \int_{u-d-\nu}^{u+d-\nu} p_{e_i^{In}}(\tau) \frac{1}{2d} d\tau d \nu .
	\end{aligned}
	\end{equation}
	According to the distribution constraint, we further obtain
	\begin{equation}
		\begin{aligned}
			& p_{\bar{o}_i^{In}}((2m\!-\!\sqrt{M}\!-\!1)d)=  \\
			&  \int_{-\sqrt{M}d}^{\sqrt{M}d} \int^{\nu+d}_{\nu-d}\!p_{s^{In}_i}\!(\mu) d\mu\! \int_{(2m\!-\!\sqrt{M}\!-\!2)d-\nu}^{(2m\!-\!\sqrt{M}\!)d-\nu} \!
			p_{e_i^{In}}(\tau) \frac{1}{4d^2}  d\tau d \nu. 
		\end{aligned}
	\end{equation}
	
\section{Proof of the Equivalence}
Assuming that 
\begin{equation}
	g(\nu, M)\triangleq \int^{\nu+d}_{\nu-d}\!p_{s^{In}_i}\!(\mu) d\mu\! \int_{(2m\!-\!\sqrt{M}\!-\!2)d-\nu}^{(2m\!-\!\sqrt{M}\!)d-\nu} \!
	p_{e_i^{In}}(\tau) \frac{1}{4d^2}  d\tau,
\end{equation}
and thus 
\begin{equation}\label{eq1}
	\begin{aligned}
	p_{\bar{o}_i^{In}}((2m\!-\!\sqrt{M}\!-\!1)d) &=  \int_{-\sqrt{M}d}^{\sqrt{M}d} g(\nu,M) d \nu.  \\
	\end{aligned}
\end{equation}
It is clear that 
\begin{equation}\label{eq2}
	\begin{aligned}
		& P(\bar{s}_i^{In}=(2m\!-\!\sqrt{M}\!-\!1)d) \\
		& = \sum^{\sqrt{M}}_{t=1} g((2t-\sqrt{M}-1)d, M) * 4d^2 \\
		& \approx  \int_{-\sqrt{M}d}^{\sqrt{M}d} g(\nu, M) d \nu  * 2d   \\
		&=  p_{\bar{o}_i^{In}}((2m\!-\!\sqrt{M}\!-\!1)d) * 2d.
	\end{aligned}
\end{equation}
Intuitively, this approximation becomes more accurate when term $\sqrt{M}d$ has limit when $M\rightarrow \infty$. For instance, when symbol $\bar{s}_i^{In}$ is uniformly distributed and the power is constrained by $\mathcal{E}_s$, $\sqrt{M}d$ would approach $\sqrt{\frac{3 \mathcal{E}_s}{2}}$. 
According to \cite{balle_arxiv2018}, (\ref{eq2}) ensures the effectiveness of employing process (\ref{proprocess}) during training. 

\end{appendices}

	\bibliographystyle{IEEEtran}
	\bibliography{IEEEabrv,Reference}
	
\end{document}